\documentclass[review]{elsarticle}

\usepackage{hyperref}
\usepackage{color}
\usepackage{subfig}

\journal{arXiv}

\begin{document}

\begin{frontmatter}

\title{Jet resonance in truncated ideally contoured nozzle}

\author{Florian Bakulu, Guillaume Lehnasch, Vincent Jaunet, \\ Eric Goncalves da Silva, Steve Girard}
\address{ISAE-ENSMA,
	Institut PPrime, Universit\'e de Poitiers, UPR 3346 CNRS\\
	1 Avenue Clement Ader, 86000 Poitiers, France}
\fntext[myfootnote]{Corresponding author: guillaume.lehnasch@isae-ensma.fr}

%\author[mymainaddress,mysecondaryaddress]{Florian Bakulu}
%\ead[url]{www.pprime.com}

%\author[mymainaddress]{Florian Bakulu}

%\author[mymainaddress]{Guillaume Lehnasch}\corref{mycorrespondingauthor}
%\ead{guillaume.lehnasch@isae-ensma.fr}

%\author[mymainaddress]{Vincent Jaunet}

%\author[mymainaddress]{\\Eric Goncalves da Silva}

%\author[mymainaddress]{Steve Girard}

%\address[mymainaddress]{ISAE-ENSMA,
%	Institut PPrime, Universit\'e de Poitiers, UPR 3346 CNRS\\
%	1 Avenue Clement Ader, 86360 Chasseneuil-du-Poitou, France}
%\address[mysecondaryaddress]{CNES,    Direction des Lanceurs, 
%	Centre National d'Etudes Spatiales (CNES)\\
%	52 rue Hillairet, Paris, 75012, France}

\begin{abstract}
Unsteady side-loads observed in supersonic nozzles operating in over-expanded regime are most often associated to intrinsic unsteadiness of the shock system and separation line, featuring random motions with mainly broadband low-frequency contributions. A tonal flow behavior, rather associated to energy peaks of fluctuating wall pressure in the middle frequency range is also found to emerge for particular  operating conditions in a Truncated-Ideally Contoured (TIC) nozzle. The corresponding flow field is here investigated to understand its origin and show how it modifies side-load properties.
The temporal and spatial organization of wall pressure and jet velocity field are first experimentally characterized based on synchronized acquisition of both wall-pressure along rings of pressure probes located within the nozzle and high-rate time-resolved PIV velocity fields measured in a plane section crossing the jet downstream of the nozzle exit. The external jet aerodynamics and internal wall pressure field are first shown to be clearly linked, but only at this frequency peak for which a significant coherence emerges between first azimuthal mode of fluctuating wall pressure and first azimuthal mode of fluctuating external velocity field.
A Delayed Detached Eddy Simulation is carried out and validated against experimental results in order to reproduce this tonal flow dynamics. The analysis of simulation data shows that the tonal flow behaviour of first azimuthal mode is indeed more largely felt within the whole flow structure where both upstream and downstream propagating waves are shown to co-exist, even far downstream of the nozzle exit. The analysis shows that both waves possess support in the jet core and have a non negligible pressure signature in the separated region. The Spectral Proper Orthogonal Decomposition of fluctuating pressure field at this tonal frequency reveals that the nature and intensity of lateral pressure forces is directed by the resonance related to the upstream- and downstream-propagating coherent structures, which imposes the shock-waves network to respond and modulate the pressure levels on the nozzle internal surface.

\end{abstract}


\end{frontmatter}

%\linenumbers

%%%%%%%%%%%%%%%%%%%%%%%%%%%%%%%%%%%%%%%%%%
%%%%%%%%%%%%%%%%%%%%%%%%%%%%%%%%%%%%%%%%%%
\section{Introduction}

The efficiency of supersonic nozzles used in space industry is based above all on the maximization of the specific impulse. This leads to consider nozzle geometries for main engine delivering an optimal thrust for relatively high altitude where pressure levels are significantly lower than sea-level conditions. Such a nozzle thus necessarily operates in over-expanded regime during the start-up of the engine at low altitude levels. In this case, the adverse pressure gradient issued from the pressure mismatch at the nozzle outlet causes a contraction of the jet column and leads to the formation of recompression shocks and flow separation inside the nozzle. Non-axisymmetric fluctuating wall pressure fluctuations are unavoidably generated in this situation. If such wall pressure fluctuations yield sufficiently high amplitude levels while remaining sufficiently coherent along the nozzle, intense side-loads may be produced and compromise the integrity of the nozzle structure.
The prediction of side-loads properties thus represents a critical issue to address in view of improving both performance and safety of space launchers.
 A comprehensive physical understanding and detailed modelling framework for side-loads are yet still clearly lacking.
 
 Side-loads features highly depend on Nozzle Pressure Ratio (NPR), nozzle geometry and external environment effects. 
These parameters first determine the general flow structure.
In particular, Truncated Ideal Contour (TIC) nozzles only show Free Shock Separation (FSS) regime. In this case, the boundary layer separates within the nozzle and a large open separation zone is formed through which the external flow is sucked down into the nozzle along the wall before being swallowed by the separated jet. The sudden deflection of the upstream flow within the nozzle is associated to the formation of a separation shock, generally connected to a Mach disk and a reflected shock. Downstream of this shock structure, the jet looks like an annular supersonic layer surrounding a subsonic core flow (downstream of the Mach disk) and surrounded by the counterflowing separation region. 
Thrust Optimized Contour (TOC) or Thrust Optimized Parabolic (TOP) nozzles feature a higher initial divergent angle, leading to the formation of a so-called internal shock just downstream of the nozzle throat, which makes the subsequent structure more complex upstream of the separation region. The FSS regime is also observed for low NPR values in TOC or TOP nozzles. However, for higher NPR values, these nozzles also exhibit a Restricted Separation Shock (RSS) regime. In this last case, the supersonic annular region is stuck to the wall again downstream of the first separation line and possibly features several successive small closed recirculation regions. A so-called \textit{cap shock} pattern is formed with a far larger Mach disk while the reflected shock now interacts with the separation shock, leading to a supersonic annular region developing closer to the wall.

As a function of the flow regime and nozzle geometry, various sources of unsteadiness can predominate.
 Among these various possibles sources, the shock/boundary layer interaction (SWBLI) has led to many studies aimed at identifying the (upstream or downstream) possible mechanisms behind the generation of local unsteady motions of the recirculation bubble and associated shock system which are formed locally in this case (see \citet{clemens:2014} for a recent review). For sufficiently strong shocks, canonical SWBLI cases feature the local formation of a closed separation region, extending over long distances of several boundary layer thicknesses and breathing in a range of frequencies which are particularly low with respect to the characteristic high-frequency range characterizing supersonic boundary layers upstream of the separation. By analogy, the asymmetry of the separation line and separation shock observed in supersonic nozzles is often considered as possibly resulting from such an intrinsic local source of unsteadiness. Most simplified models for side-loads, as proposed by \citet{Schmucker1973-2} or \citet{Dumnov1996} are, indeed, based only on the consideration of such local oscillations of the separation shock, driven by local flow properties around the shock.
Unsteady numerical simulations of such complex flow fields have also clearly revealed the fundamental role of convective instabilities developing in the jet shear layer \cite{deck2002numerical}, contributing to most part of the higher frequency contributions to wall pressure fluctuations. It is worth noting that the whole shock structure and separation line gradually shift downstream for increasing NPR so that the initial conditions driving the conditions of mixing layer development change. As a function of NPR, various types of dominant shear instabilities could be expected. Preliminary observations of such variable instabilities have been numerically observed in FSS regime in a TOC nozzle \cite{shams2011} yet without showing strong evidence of significant change of the side-loads behavior in this case.
In addition to these probably most common and dominant mechanisms into play, other phenomena have been shown to play a non-negligible role in more particular situations.
For example, following \citet{stark2009experimental}, the asymmetry of the separation/shock system might sometimes be related to the non-homogeneity of the upstream laminar to turbulent boundary layer transition at relatively low NPR (thus also corresponding to low Reynolds numbers). 
In FSS regime, for increasing NPR, the global structure shifts downstream and appears more radially extended. The greater proximity of the shear layer to the nozzle wall observed for certain values  of NPR indeed allows enhanced levels of seeding of pressure perturbations coming from the shear layer into the recirculation region and even sometimes lead to random intermittent impingement of the separeted shear layer on the nozzle wall \cite{Verma14}. 
For separation point sufficiently close to the nozzle exit, the shape of the recirculation region tends to change from a rather cylindrical shape to a conical one. An asymmetric change can exacerbate the  fluctuations of momentum difference between the core flow downstream of the Mach disk and the surrounding flow downstream of the separation shock, thus leading to enhanced flow unsteadiness \cite{verma2017}.
As a function of the exact geometry of the nozzle lip and external environment configuration driving the coflowing conditions, the formation of a small secondary re-circulation zone can also be observed within the nozzle close to the nozzle exit \cite{Hadjadj15}. This may contribute to modulate the forcing of external pressure fluctuations through the separation zone due to the proximity of jet mixing layers to the wall \cite{georges2014influence}.
In TOC nozzles featuring RSS regime, a far larger Mach disk is produced with more pronounced curvature levels, leading to the generation of a large recirculating region in the subsonic core of the jet downstream of a this Mach disk. This large recirculation bubble presents an intrinsic complex three-dimensional dynamics \cite{shams2013unsteadiness}. For this regime, the position of the restricted separation regions close to the wall move downstream as the NPR increases, so that they can intermittently open to the ambient atmosphere at particular NPR values. This so-called "end-effect regime" is known to be associated to particularly significant levels of unsteadiness and the generation of particularly intense lateral forces \cite{Nguyen2003} \cite{deck2009delayed}.
The integral of the pressure forces resulting from the contribution of all these potential sources of instability then generally presents
a random character, mostly dominated by low-frequency side-loads activity \cite{deck2002numerical}\cite{shams2013unsteadiness}.

The present study more particularly focuses on the unsteady mechanisms encountered in FSS regime in presence of tonal flow behavior. It aims at identifying the hidden global flow structure responsible for this particular behavior and inferring its potential consequences in the generation of lateral aerodynamic forces.  A TIC nozzle is considered, with a full flowing design Mach number equal to $M_d=3.5$ and some operating conditions corresponding to an equivalent perfectly expanded jet Mach number $M_j=2.09$. The non-dimensionalization retained for this study is based on the nominal conditions. A given nozzle pressure ratio is set from the prescription of an upstream total pressure $P_j$ and fixed external quiescent atmosphere in standard conditions. The NPR and the nozzle geometry thus determine the equivalent perfectly expanded Mach number $M_j$, exit jet velocity $U_j$ and jet diameter $D_j$ of this equivalent perfectly expanded jet at Mach $M_j$.
The particular evolution of spatio-temporal structure of internal wall pressure field in the present TIC nozzle has been described in detail in \citet{jaunet2017wall} in a large range of NPR values. In addition to expected broadband low-frequency contributions due to shock/separation line movement and high-frequency contributions easily associated to advection of coherent structures in the mixing layer, discrete energy peaks of high amplitude have been identified in the intermediate frequency range. Through the use of rings of Kulite sensors, allowing azimuthal decomposition of pressure fluctuations, it has been shown that each peak corresponds to the activity of a particular azimuthal mode. The peak of highest amplitude corresponds to the first non-symmetric azimuthal mode $m=1$ which is of most interest for its unique role in the possible generation of side-loads \cite{Dumnov1996}.
In a relatively narrow range of operating conditions,  whatever the probe location considered in the streamwise direction, the amplitude of the energy peak has been shown to emerge far more largely in a particularly narrow range of operating conditions, around $M_j=2.09$, and at a Strouhal number $St=fD_j/U_j=0.2$, where $f$ stands for the frequency. This working condition is here retained accordingly for the present study to focus on the tonal dynamical behavior of TIC nozzle flow. 

It should be recalled that the emergence of that kind of discrete acoustical tone in nozzles has already been related in various studies to transonic resonance. It appears to be more often observed in nozzles yielding relatively low area ratio, like in conical nozzles with low divergent angle for \citet{zaman2002investigation}. The emergence of a similar acoustic tone has yet also been reported for a TOP nozzle featuring a far higher exit to throat area ratio, at high NPR in RSS regime  \cite{donald2014sound}, leading the authors to speculate that a transonic resonance or a screech loop could be present. 
This mechanism of transonic resonance indeed involves the formation of a standing pressure wave between the nozzle exit and separation shock \cite{zaman2002investigation} corresponding to a well-defined feedback loop established through perturbations traveling downstream in the shear layer from the interaction zone between the separation shock/ boundary layer interaction zone and upstream propagating waves traveling in the central subsonic zone downstream of the shock \cite{olson2013mechanism}. A stagging behavior is likely to be observed, with a switch from high-frequency odd-harmonic modes to the lower frequency fundamental mode associated with the distance between the shock and the nozzle exit.
An important aspect is however that the mechanism appears to be essentially axisymmetric and only the behavior of the axisymmetric mode was experimentally found to be affected by the transonic resonance. In addition, \citet{larusson2017dynamic} have carried out a Dynamic Mode Decomposition (DMD) analysis, just based on snapshots obtained from perturbed URANS computations in axisymmetric formulation. These authors have shown that it can already enable the identification of dominant modes and frequencies in fair agreement with the standing wave experimentally observed \cite{loh2002numerical}. Such observations thus support the idea that this mechanism could probably not be the main responsible for tonal behavior of non-symmetric mode and may not be expected to contribute directly to side-loads generation.
Following a previous analysis carried out on the present TIC nozzle \cite{jaunet2017wall}, it is indeed more clearly demonstrated that the peak frequency can not be simply explained by transonic resonance. Due to the shift of the whole shock structure towards the nozzle exit when NPR is increased, transonic resonnance mechanism is indeed naturally expected to produce resonances at higher frequencies when NPR is increased. 
However, by tracking the peak frequency of internal wall pressure in a relatively wide range of NPR where this peak could be observed (even with lower amplitude than the amplitude observed at $M_j=2.09$), the evolution of the frequency peak was shown to follow an opposite trend. In addition, no staging behavior could be observed in that case.
The trend of the frequency evolution with respect to the NPR was indeed rather recalling a screech behavior \cite{tam1986proposed} (with a decrease of this frequency with $M_j$). Surprisingly, it has not been possible to identify any tonal component in the radiated sound during the experiment, like it should be the case in presence of screech. A possible \textit{pseudo-screech} mechanism, more related to the internal subsonic core of jet flow (with a possible masking effect by the surrounding supersonic shear layers) has been suggested accordingly in \cite{jaunet2017wall}. 
As expected, internal wall pressure fluctuations were also found to have mainly positive phase velocities, thus corresponding to perturbations advected in the downstream direction. It was also checked that these internal fluctuations were largely uncorrelated with external velocity fluctuations for most frequencies, due to the rapid development of turbulence. However, at the particular tone frequency $St=0.2$, the mode $m=1$ exhibited a negative phase velocity of wall pressure fluctuations while the amplitude of the transfer function between internal pressure functions and external velocity fluctuations has been shown to be significantly increasing. This study thus suggested for the first time a possible synchronization of upstream- and downstream-propagating waves.
This idea has partially been supported by some recent numerical DDES observations of the flow in another TIC nozzle featuring a similar tonal behavior, at intermediate frequency peak, associated with non-symmetric pressure mode by \citet{martelli2020flowdynamics}. In this study, the authors propose a possible loop scenario, in which the intermittent passage of turbulent structures of the detached shear layer interact with the triple point, causing a significant distortion of the Mach disk and the formation of intense vortex shedding in the central subsonic core of the jet. The interaction of these vortical structures advected downstream with the secondary shock structure would lead to the emission of acoustic waves travelling back upstream through the outer subsonic region up to the separation line to trigger new shear layer instabilities

The present paper aims at reviewing some recent studies carried out to further investigate this jet configuration and the resonance mechanism observed in the present TIC nozzle at $M_j=2.09$ for $St=0.2$.  An experimental set-up has been designed in order to assess more directly the effective correlation between internal and external azimuthal modes of fluctuating fields. The whole set of experimental data have allowed a fine tuning of a Delayed Detached Eddy Simulation (DDES) then used to reproduce the phenomenon and educe the coherent content related to the resonance loop. 

The paper first describes the experimental set-up and the main ingredients of the simulation tools in section \ref{investigation-tools}. The main features of the average flow and spatio-temporal organization of fluctuations are summarized in section \ref{global-flow-organization}. The coherent structure eduction through  Spectral Proper Orthogonal Decomposition (SPOD) method and the link of this coherent structure with the generation of lateral forces are then presented in section \ref{global-coherent-structure} and \ref{lateral-pressure-forces} before summarizing the main conclusions of the study in section \ref{conclusions}.

%%%%%%%%%%%%%%%%%%%%%%%%%%%%%%%%%%%%%%%%%%
%%%%%%%%%%%%%%%%%%%%%%%%%%%%%%%%%%%%%%%%%%
\section{Investigation tools}
\label{investigation-tools}

%%%%%%%%%%%%%%%%%%%%%%%%%%%%%%%%%%%%%%%%%%
\subsection{Experimental set-up}

The experimental campaign is conducted in the S150 supersonic wind tunnel at Pprime Institute. A rigid sub-scale TIC nozzle
is considered  with a divergent length $L = 0.1827$ $m$, a throat diameter $D_t=0.038$ $m$ and an outlet diameter $D = 0.097$ $m$, which lead to a full-flowing flow condition with a Mach number $M=3.5$. 
Its geometry has been designed by combining the standard Method of Characteristics (MOC) in axisymmetric formulation and a correction to account for the boundary layer development estimated by an integral approach.
The nozzle is supplied with cold (total temperature around $260$ $K$) and desiccated high-pressure air flow with low turbulence level to reach a condition of NPR corresponding to a fully expanded Mach number $M_j=2.09$. In the following, the fully expanded jet velocity $U_j$ and jet diameter $D_j$ corresponding to this value of $M_j$ are used to define a non-dimensional time $t^*= t U_j / D_j$ and a Strouhal number $St=f D_j/U_j$ based on time $t$ or frequency $f$ respectively.

\begin{figure}[htbp]
	\centering
\includegraphics[trim = 2cm 5cm 2cm 2cm, clip, width=0.7\textwidth]{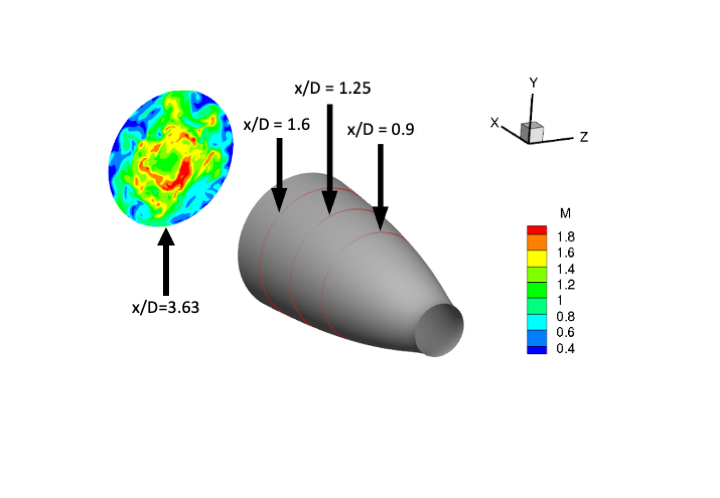}
\caption{Positions of rings of wall pressure Kulite sensors and PIV plane.}
	\label{figure1}
	\end{figure}

Synchronized time-resolved stereo-PIV and wall pressure measurements are carried out. Wall pressure fluctuations are acquired using sensors placed inside the nozzle whereas velocity samples are obtained in an external plane orthogonal to the streamwise direction. The measurement locations are illustrated in Fig.~\ref{figure1}.
The nozzle is equipped with 18 flush-mounted Kulite XCQ-062 pressure transducers, distributed along 3 rings of 6 transducers placed equidistantly along the circumference. The first ring is located at $x/D=0.90$ where $x$ is the axial distance from the nozzle throat. It is thus located close to the separation shock occurring here at the middle of the nozzle in this case (whose length is about $1.8D$). The two other rings are in the recirculation zone at $x/D=1.25$ and $x/D=1.60$. The pressure is measured during $10^{5} t^*$ with a sample rate  corresponding to $St =  20$. Note that a low-pass filtering has also been applied on wall pressure signals to fit the lower resolution limit of the PIV system before computing the correlation between the pressure and velocity signals.

Stereo-PIV measurements are carried out with a diode pumped 527nm 30mJ Continuum MESA-PIV laser and  high repetition rate Photron cameras, synchronized with the wall pressure acquisition system. The flow was seeded
using $SiO_2$ particles whose mean diameter
has been estimated to $0.3\mu m$ and their relaxation time to $0.019 ms$
\cite{lammari:1996}. This is sufficient for the time scales of interest in this
paper. The flow was seeded internally, via a seeding cane placed inside the resting
chamber, and externally, with a seeding cane aligned with the jet axis. Note that it was checked that the flow was not affected by the seeding system
  by comparing wall pressure measurements with and without the seeding canes.
The velocity data analyzed for this study are extracted in a plane normal to the jet axis and located at $x/D=3.63$ with a Field of View of about two nozzle diameters.
This particular position of the measurement plane is chosen \textit{a priori} in order to favor the detection of any possible link between internal and external fluctuations which is likely to be rapidly masked by the development of turbulence in mixing layers. In the present case, it nearly coincides with the end of the pseudo-potential core of the supersonic jet and corresponds to the position where the maximal amplitude of the transfer function between internal and external fluctuating fields has previously been detected \cite{jaunet2017wall}. 
The initial post-processing of PIV images (of size equal to $1024$ by $1024$ pixels) was done with decreasing window size from $128 \times 128$ to $32 \times 32$ pixels. A total of three passes with a $50\%$ window overlap was used. The first pass was performed using square windows, adaptive PIV algorithm \cite{Scarano2002} was used thereafter.
Vectors are validated using a universal outlier detection \cite{Westerweel2005} together with the standard correlation peak-ratio criterion. Only the validated vectors were used in the following processing stages while the other wrong vectors were flagged. A few images unavoidably contained few spurious vectors, due to the difficulties to ensure a perfectly homogeneous seeding during the whole time sequence of data acquisition. In the present case, the erroneous vectors yet remained small enough in number and sufficiently dispersed in the images to consider an \textit{a posteriori} spatial data reconstruction. A spatial interpolation ($3^{rd}$ order Lagrange polynomials) has thus been used to rebuilt the local lacking information.
Around $21000$ successive PIV images have been retained for the present case at $M_j=2.09$, corresponding to a period of  $2.1 \times 10^{4} t*$ with a sampling rate of $St = 1$. After the application of the PIV correlation algorithms, Lagrange interpolation has also been applied to interpolate the velocity data on a cylindrical grid suitable for the extraction of azimuthal Fourier modes of velocity fluctuations. Note that it has been verified \textit{a priori}, based on other sets of perfectly reliable reference velocity data, that this whole numerical treatment only marginally bias the spectral content of the azimuthal modes of interest in the spectral range considered in the study.

%\textcolor{green}{
%The obtained Field Of View (FOV) for PIV planes is about 2 nozzle diameters in the $(Y,Z)-plane$ (cross planes normal to the jet axis, see figure \ref{figure1}). The PIV was ran at a sampling rate of 20 KHz and the flow was seeded using $SiO_2$ particles whose mean diameter has been estimated to $0.3\mu m$ and their relaxation time to $0.019 ms$ \cite{lammari:1996}. This is sufficient for the time scales of interest in this paper. The flow was seeded internally, via a seeding cane placed inside the resting chamber, and externally, with a seeding cane aligned with the laser plane. It was checked that the flow was not affected by the seeding system by comparing wall pressure measurements with and without the seeding canes. The initial post-processing was done with decreasing window size $128 \times 128$ to $32 \times 32$. A total of three passes with a $50\%$ window overlap was used. The first pass was performed using square windows, adaptive PIV algorithm \cite{Scarano2002} was used thereafter. Vectors are validated using a universal outlier detection \cite{Westerweel2005} together with the standard correlation peak-ratio criterion. Only the validated vectors were used in the following processing stages. The vector fields were filtered to discard the faulty images due to inhomogeneous seeding resulting in an amount of $7800$ vector fields. They represent a sufficient number of independent snapshots to converge the relevant statistical quantities studied.\\}
 
%%%%%%%%%%%%%%%%%%%%%%%%%%%%%%%%%%%%%%%%%%%%%%%%%%%%
%%%%%%%%%%%%%%%%%%%%%%%%%%%%%%%%%%%%%%%%%%%%
\subsection{Numerical set-up}

The over-expanded jet in the present TIC geometry is numerically reproduced by carrying out a DDES with the in-house code \textit{PHOENIX} developed at \textit{Pprime Institute}. 
Simulation of unsteady flow features of such supersonic nozzle flows is particularly challenging. A sufficiently high resolution is required both near walls to capture attached boundary layers and further downstream when unsteady turbulent structures develop. This leads to very restrictive time steps while a long simulation time is required to capture the expected low-frequency features of integrated wall pressure forces. 
While this kind of nozzle flow configuration has been widely studied with RANS simulations, it is worth recalling that, until very recently, only very few unsteady simulations with hybrid approaches, similar to the one adopted in the present study, have been attempted. Such unsteady simulations have been carried out in the case of planar nozzle configuration \cite{martelli2019characterization} or truncated ideally contoured nozzles \cite{deck2009delayed} \cite{shams2013unsteadiness}.
The first hybrid simulations of TIC nozzle flow have more recently been reported in \cite{goncalves2017hybrid} and then in \cite{martelli2020flowdynamics}.

\begin{figure}[htbp]
	\centering
	\begin{minipage}[t]{5cm}
	\includegraphics[width=4cm, trim={2cm 0cm 2cm 0.5cm}, clip]{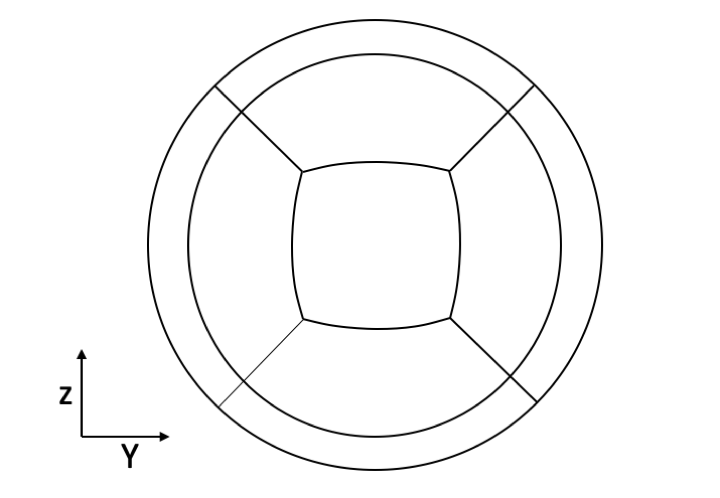}
	 \includegraphics[width=4cm, trim={2cm 0cm 2cm 0.5cm}, clip]{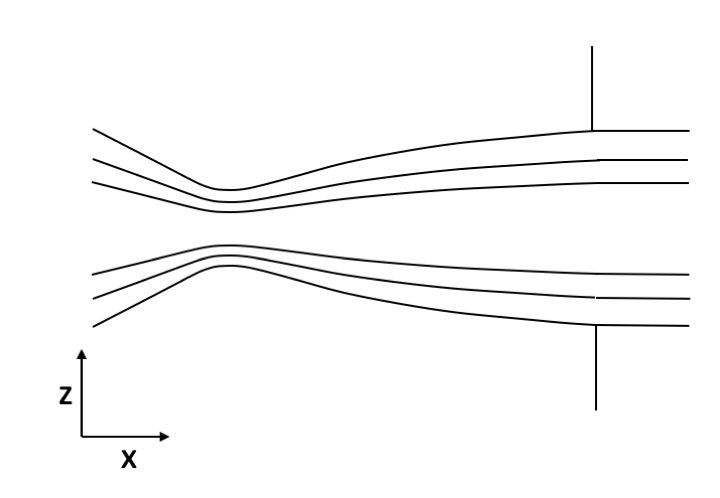}
	 \end{minipage}
	 \begin{minipage}[c]{6cm}
	\includegraphics[width=6cm, angle=90, trim={1cm 12cm 3cm 0cm}, clip]{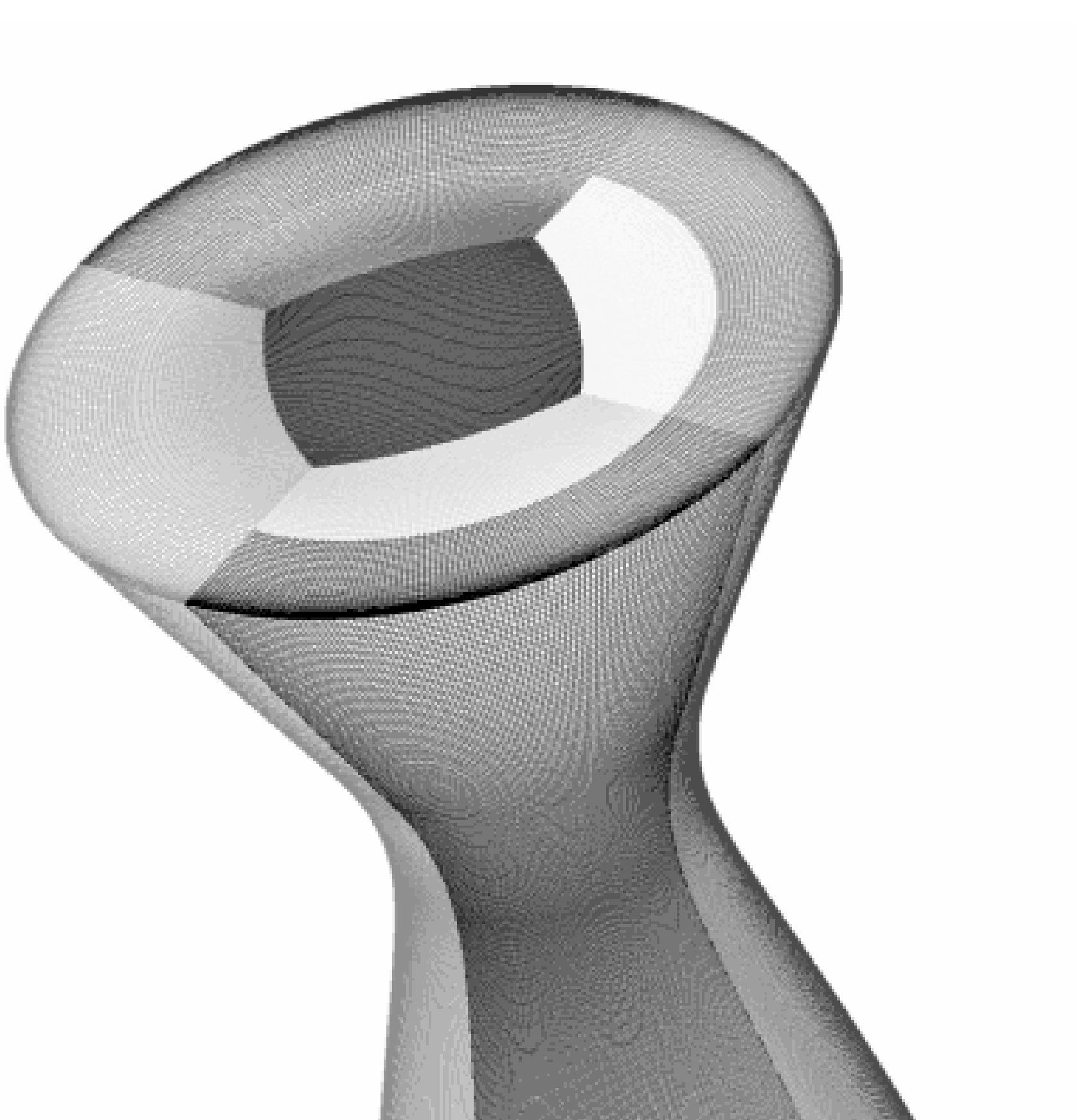}
	\end{minipage}
	\caption{Sketch of the computational multiblock topology in cross plane (top left) and streamwise plane zoom in nozzle region) (bottom left) and perspective view of nozzle mesh in the convergent region.}
	\label{figure2}
	\end{figure}

The main numerical ingredients used for the present study are summarized as follows.
%%%%%%%%%%%%%%%%%%%%%%%%%%%%%%%%%%%%%%%%%%%%%%%%%%%%
% Computational domain and mesh
The whole internal TIC geometry considered for the experiment (including the end of the convergent part of the nozzle) is meshed with a 5-blocks butterfly mesh topology for the internal part of the flowfield, surrounded by an additional ring of 4 blocks added to improve the control of the distribution of mesh points near the wall. The resulting mesh topology is illustrated in Fig.~\ref{figure2}. This meshing strategy \textit{a priori} allows a better control of mesh sizes distribution compared to the case where two-dimensional meshes are simply extruded in the azimuthal direction. In this last case, the exaggerated disparity in mesh sizes in the azimuthal direction between the axis and the nozzle wall may lead to artificial numerical oscillations or artifacts, in particular close to the axis. 
 The last slice of the mesh at the nozzle exit is extruded in the streamwise direction over a distance of $28D$ in the external domain by applying a rapid stretching. This central mesh region downstream of the nozzle exit is surrounded by an additional 4 blocks annular mesh layer extending up to $5D$ in the radial direction. Non-reflective open boundary conditions are applied at these far-field boundaries. Note that a wall condition is also applied around the nozzle at the level of the jet exit plane. Despite it can not allow an accurate representation of the slight coflow unavoidably met during the experiment where this plane is not present, this choice is retained in order to limit the cost of the simulation while maintaining a reasonable representativity of external flow conditions. The global mesh includes $795$ points in the streamwise direction, $159$ points in the radial direction and $400$ points in the azimuthal direction, leading to a total of around $49$ million cells. 
 In order to tackle the constraints met for a too much refined grid near the wall, wall functions are applied according to the formulation described in \cite{Goncalves01}. The mesh resolution has been found to be satisfactory from \textit{a posteriori} analysis of numerical results. In particular, the $y^+$ values in the adjacent cells to walls vary between 10 and 15 in regions where boundary layers are still attached. It has also been checked that the mesh resolution globally satisfies LES requirements in the separated jet region. The ratio of subgrid viscosity over molecular viscosity reaches values less than $10$ to $20$ in the whole central part of the jet (yet still largely unaffected by the spatially developing shear turbulence) and typically lies in the range $[80:100]$ within the supersonic shear layers at least up to the region surrounding the secondary Mach disk downstream of the nozzle exit. Beyond $x/D=3$, the consideration of a rapid mesh coarsening rapidly leads to higher values, which limits the relevance of the smallest scales observed in the downstream in LES mode.

%%%%%%%%%%%%%%%%%%%%%%%%%%%%%%%%%%%%%%%%%%%%%%%%%%%%
% DDES model
The present DDES approach is based on Spalart-Allmaras model and is classically built by allowing the length scale in the model to be
proportional to a representative scale related to the grid size far from the wall, while remaining given by the RANS model for small distances from the wall \cite{Spalart06}. Such models based on Spalart-Allmaras models are often adopted in studies of similar nozzle flow configurations and, if correctly tuned, they have already proved to lead to satisfactory results for nozzle flows  \cite{deck2009delayed,goncalves2017hybrid,martelli2019characterization}.

The equation for the pseudo viscosity $\tilde{\nu}$ of the model reads:
\newcommand{\der}[2]{\frac{\partial #1}{\partial #2}}
\begin{equation}
\der{\rho \widetilde{\nu}}{t} + \der{}{x_l} \left[\rho \,u_l \,
\widetilde{\nu} - \frac{1}{\sigma} \left(\mu+\rho \widetilde{\nu}
\right) \der{\widetilde{\nu}}{x_l} \right] \,=\, c_{b1}
\widetilde{S} \, \rho \, \widetilde{\nu} +
\frac{c_{b2}}{\sigma} \der{\rho \widetilde{\nu}}{x_l}
\der{ \widetilde{\nu} }{x_l} - c_{\omega 1}f_{\omega} \, \rho \frac{{\widetilde{\nu}}^{2}}{\tilde{d}^{2}}
\end{equation}
where $\widetilde{S}$ is a modified vorticity magnitude,
$f_{\omega}$ is a near-wall damping function and $\sigma$, $c_{b1}$, $c_{b2}$, $c_{\omega 1}$ are the model constants. The eddy viscosity is defined as $\nu_t= f_{v1}
\tilde{\nu}$ where $f_{v1}$ is a correction
function designed to guarantee the correct boundary-layer behavior in the near-wall region.
The new distance to the wall $\tilde{d}$ used in the model is defined as:
\begin{eqnarray}
\tilde{d} &=& d - f_d \ \hbox{max}(0,d-C_{DES} \Delta)
\end{eqnarray}
The introduction of the function $f_d$ ensures that RANS mode is enforced in near-wall region. This delay of the transition between RANS and LES mode aims at preventing the phenomena of model stress depletion and possible grid-induced separation, due to excessive reduction of the eddy viscosity in the region of switch (grey area) between RANS and LES modes. It is defined as:
\begin{eqnarray}
f_d &=& 1-\hbox{tanh}\left(\left[8r_d\right]^{3}\right) \quad
\hbox{with} \quad r_d = \frac{\tilde{\nu}}{\kappa^{2} d^2 \sqrt{U_{i,j}U{_{i,j}}}}
\end{eqnarray}
where $U_{i,j}$ is the velocity gradient, $\kappa$ the von Karman constant and $d$ the effective distance to the wall. The constant $C_{DES}$ has been set to its reference value $0.65$.

A key ingredient of the present methodology is the use of a hybrid characteristic sub-grid length scale $\Delta$:
\begin{equation}
\Delta = \frac{1}{2} \left[\left(1-\frac{f_d-f_{d0}}{\mid f_d -f_{d0} \mid}\right) \Delta_{max}+ \left(1+\frac{f_d-f_{d0}}{\mid f_d -f_{d0} \mid}\right) \Delta_{vort} \right]
\end{equation}
This scale depends on the flow itself through the function $f_d$, and is based on a blend of the usual characteristic length $\Delta_{max}=\hbox{max}(\Delta_x,\Delta_y,\Delta_z)$  enforced in boundary layers and another vorticity-based scale $\Delta_{vort}$ which depends on the local flow properties. This last scale is defined according to:
\begin{equation}
\Delta_{vort}=\sqrt{\frac{\sum_i \mid \omega.S_i \mid}{2 \mid \omega \mid}}
\end{equation}
where $\omega$ is the vorticity vector, $S_i$ the oriented surface $i$ of a given cell. It takes into account the direction of the vorticity vector in order to reduce the issue of delayed development of convective instabilities in mixing layers, in particular due to strongly anisotropic cells \cite{chauvet2007zonal,deck2012recent}. The weighting parameter $f_{d0}$ requires a specific calibration as a function of the flow considered. For the present study, a value of $0.94$ has been found to lead to satisfactory results.

%%%%%%%%%%%%%%%%%%%%%%%%%%%%%%%%%%%%%%%%%%%%%%%%%%%%
% Numerical solver
The hybrid RANS/LES equations are integrated with a finite-volume discretization. The convective flux of main conservative variables at cell interfaces is computed with the Jameson-Schmidt-Turkel scheme \cite{jameson1981numerical} for which the dispersive error is canceled. 
It is based on the addition of artificial viscosity through both a second-order dissipation term $D_2$ and a fourth-order dissipation term $D_4$, whose values are given in Table~\ref{tab.ParamNum}. The dissipation scaling factor is
replaced with an anisotropic formulation \cite{Swanson98}, which
produces a significant improvement in accuracy for high-aspect-ratio meshes.
The two corresponding parameters have been tuned during the simulation to ensure both enhanced stability during the initial numerical transient and then to reduce the numerical dissipation and obtain the best possible robustness/accuracy compromise during the phase of data acquisition for physical analysis. 

The classical upwind Roe scheme \cite{roe1981approximate} is also used for improving robustness of the evaluation of convective flux components for the turbulence
transport equations. The second-order accuracy is obtained by introducing a flux-limited dissipation \cite{tatsumi1995flux}.
The viscous terms are discretized with a second-order space-centered scheme. Moreover, weighted schemes are implemented to take into consideration the mesh deformation. The centered numerical fluxes and the gradient computations are corrected by using a weighted
discretization operator, as described in \cite{Goncalves04}.

A dual time stepping implicit method \cite{jameson1991time} is combined with an explicit third order Runge-Kutta method for time integration. 
The former method has been introduced to tackle the lack of numerical efficiency of approaches based on global time stepping. The derivative with respect to the physical time is discretized by a second-order scheme. Between each time step, the solution is advanced with a dual fictitious time and acceleration strategies developed for steady problems can be used to speed up the convergence in fictitious time. A matrix-free implicit method is considered for each sub-iteration. It consists in solving a system of equations arising from the linearization of a fully implicit scheme, at each time step. The key feature of this method is that the storage of the Jacobian matrix is completely eliminated, which leads to a low-storage algorithm. The implicit time-integration procedure leads to a system which is solved iteratively using the point Jacobi algorithm. The numerical parameters used are given in Table~\ref{tab.ParamNum}.
\begin{table}[htbp]
\centering
\begin{tabular}{lc}
%  numerical parameters                    &            \\
\hline

%dimensionless time step, $\ds \Delta t*=\Delta t \sqrt{\gamma r T_i} /D_{throat} & $2.38\,10^{-2} %t*$  \\
dual time stepping sub-iterations         &   100    \\
CFL number                                &  0.5     \\
implicit Jacobi iterations                &   14      \\
2nd and 4th order dissipation parameter   &   0.6 ; 0.012    \\
 \hline
\end{tabular}
\caption{Numerical parameters used in the DDES simulation.} \label{tab.ParamNum}
\end{table}

%%%%%%%%%%%%%%%%%%%%%%%%%%%%%%%%%%%%%%%%%%%%%%%%%%%%
% Initialization and time integration
The flow within the nozzle is initialized with quiescent conditions. In order to limit the intensity of waves initially produced at the beginning of the numerical transient, the upstream conditions of pressure first correspond to a lower NPR. The inlet pressure is then only progressively increased to reach the desired condition of NPR. The flow instabilities (in particular convective instabilities developing in mixing layers) set up naturally during the numerical transient and are maintained without need to seed artificial upstream noisy perturbations.   
During the phase of data acquisition for the analysis, the time step is reduced to sufficiently small values of $\Delta_t =0.0238 t^* $ to allow a relevant spectral analysis of numerical data. Around $2900$ snapshots of the flowfield have been generated, covering a physical time duration of $676 t^*$, and already requiring around $380000$ CPU hours on TGCC (Curie) computational center. 
This time duration is enough to obtain good convergence levels for first and second order statistics to within a few percents in the whole flowfield in the physical domain. As expected, longer time of computations (at a non-affordable cost) would still be necessary for spectral analysis, in particular to reach more converged evaluations of coherence levels between internal and external fluctuating fields. It is shown in the following that the present database is however sufficient to extract the most relevant information at the tonal frequency of most interest.

\section{Statistical description of global flow organization and validation}
\label{global-flow-organization}

\begin{figure}[htbp]
\centering
\includegraphics[width=0.75\textwidth, trim={0 0 0 0},clip]{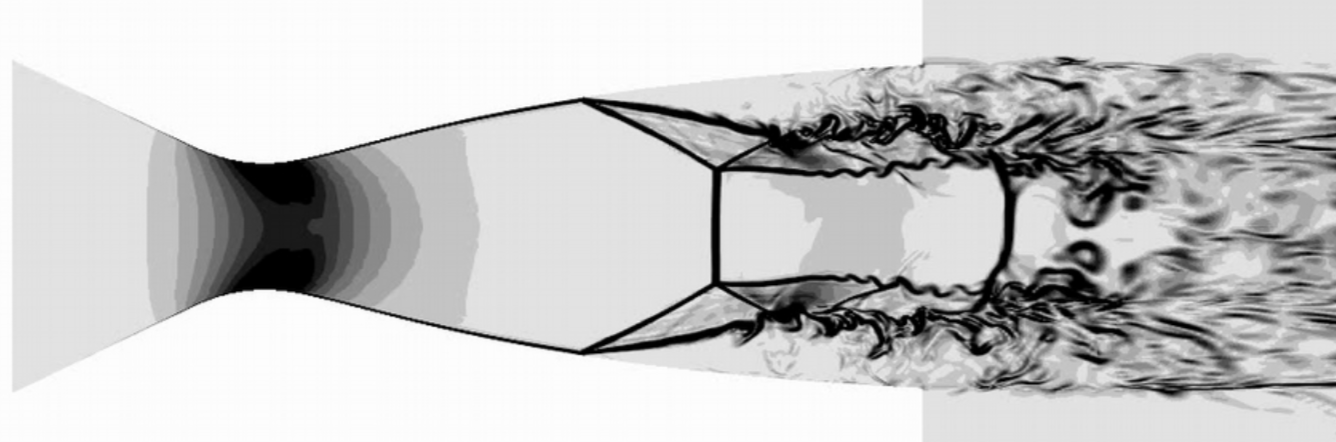}
\caption{\label{schlieren} Instantaneous pseudo-Schlieren visualization of the TIC nozzle jet at $M_j=2.09$.}
\end{figure}

\begin{figure}[htbp]
\centering
\includegraphics[width=12cm, trim=0cm 4cm 0cm 3cm, clip]{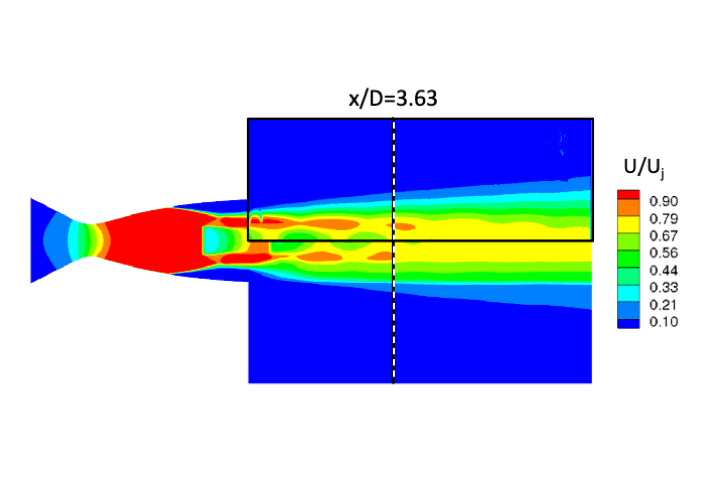}
\caption{Average streamwise velocity field: reference x-y PIV data (surrounded by black rectangular box) versus overall numerical prediction. Position of the y-z cross-PIV plane (vertical black and white dashed line). }
\label{figure-comparison-velocity}
\end{figure}

An instantaneous pseudo-Schlieren visualization is first presented in Fig.~\ref{schlieren}. It shows that the essential expected features of the flow are well reproduced.  The open separation here arises from around the middle of the nozzle divergent.
The average position of the separation line experimentally observed from pressure data (and confirmed with oil films) at $x/D=0.93$ nearly coincide with the one numerically evaluated at $x/D=0.95$ through wall pressure gradients. 
The separation produces the so-called separation shock and large Mach disk due to irregular shock reflection at the axis. This structure is followed in the downstream by a supersonic annular mixing layer surrounding a subsonic core and surrounded by the reverse recirculation region close to the wall. The flow re-acceleration in the supersonic jet then leads to the formation of a secondary Mach disk here located slightly downstream of the nozzle exit. The mean location and extent of both mixing layers and shock structures downstream of the nozzle exit are in good agreement with experimental data available, as shown in particular with the average streamwise velocity field obtained in Fig.~\ref{figure-comparison-velocity} with a comparison with some reference PIV data from a previous study \cite{jaunet2017wall}.

It is worth noting that in the present simulation, instabilities naturally develop within the annular mixing layers due to the forcing by the arising global jet oscillations. They first emerge during the numerical transient and are maintained when the flow is established. Numerical flow visualizations reveal that the shock motion is dominated by back and forth motion (mode $m=0$) with irregular precession (mode $m=1$) alternately clockwise or counter-clockwise direction.
Phases of more intense tilting and/or distortion of the shock structure is occasionaly associated to more significant amplitude of oscillation of the jet column, producing larger and more intense coherent structures travelling in the shear layer. However, the shock oscillations observed never produce a sufficiently important change of the Mach disk curvature to generate large vortices close to the axis downstream of the Mach disk as it could have been reported in \cite{martelli2020flowdynamics}.

The shear layer instabilities seem to be first slightly delayed as could be expected, as it could be yet classically observed in absence of any specific upstream forcing treatment. Downstream of the secondary Mach disk, they admittedly also suffer from excess of numerical diffusion due to the rapid mesh stretching in the downstream region, chosen to limit the computational cost. By comparison with experimental observations in Fig.~\ref{figure-comparison-velocity}, the compression/expansion regions within the annular mixing layer yet appear only slightly shifted downstream of their position experimentally observed. 
The maximal differences between numerical predictions and experimental measurements of average streamwise velocity typically also remain limited to less than $10\%$ within the mixing layer region at the location of the cross PIV plane used in the following to examine further the link between internal and external fluctuations.

\begin{figure}[htbp]
	\centering
	\includegraphics[width=3in]{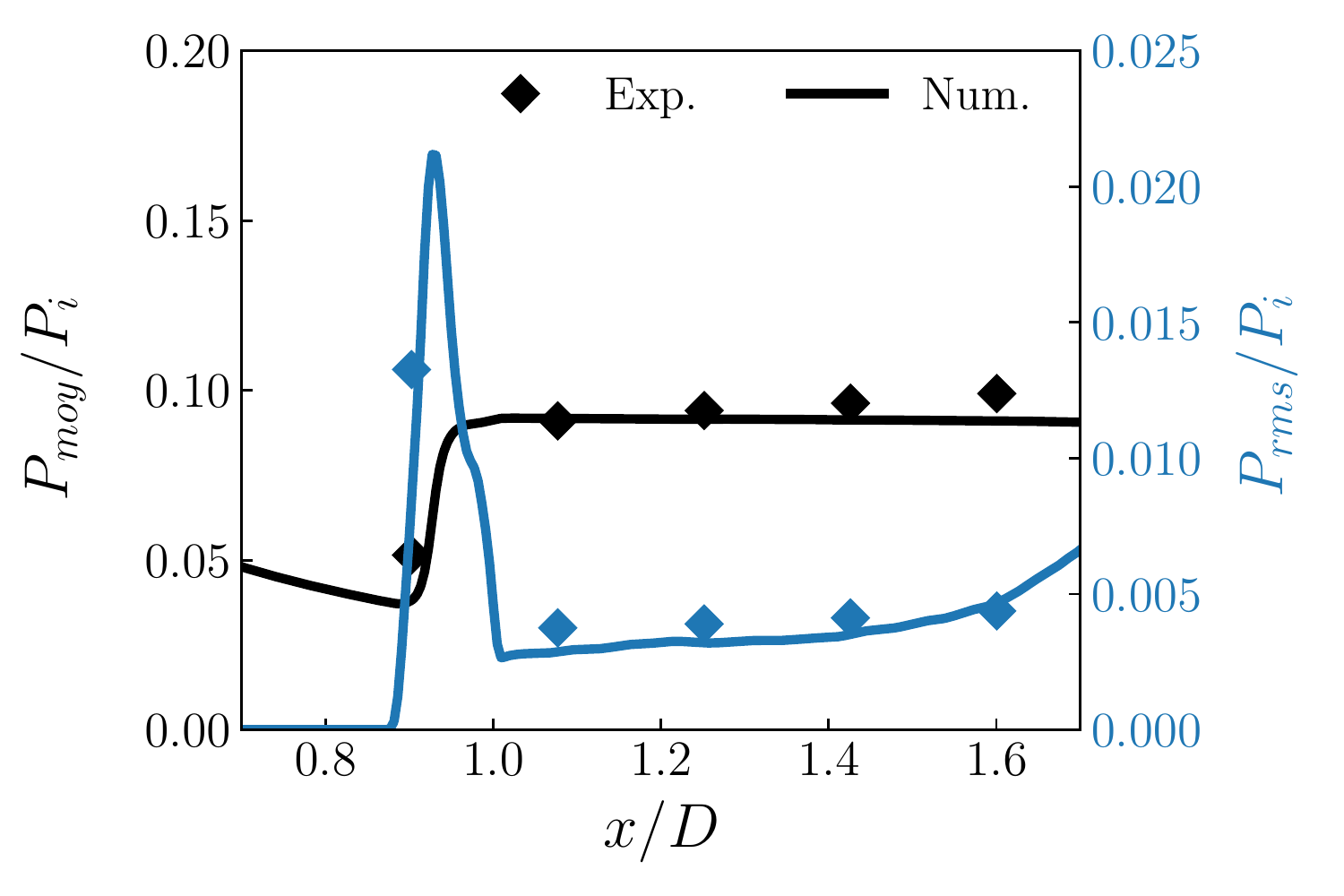}
	\caption{Average (in black) and RMS (blue) wall pressure distributions along the nozzle wall.}
	\label{figure-comparison-pressure}
	\end{figure}

A particularly good agreement is observed between numerical results and available experimental wall pressure data within the nozzle. These streamwise distributions of average and rms of wall pressure fields are presented in Fig.~\ref{figure-comparison-pressure}. They indicate that the present simulation satisfactorily captures the shock position and the jump of both average pressure level and intensity of pressure fluctuations in this critical region.

%%%%%%%
%%%%%%%
%%%%%%%

%The numerical results obtained are compared with the experimental results available for $M_j=2.09$. The mean separation line is located at $x/D=0.95$, very close to the reference experimental value $x/D=0.93$. The distributions of average and fluctuating wall pressure along the nozzle are shown in figure~\ref{figure-comparison-pressure} and the agreement between the two sets of data shows that the numerical model satisfactorily reproduced the mean flow salient features.

%The comparison of experimental (from a former PIV campaign) and numerical streamwise velocity fields in figure~\ref{figure-comparison-velocity} also shows that the expected external jet flow structure is quite well reproduced. The spreading of mixing layer downstream of the nozzle exit is yet slightly over-estimated due to the rapid mesh coarsening in the streamwise and radial directions in the external domain. This leads to a small shift of the jet flow structure (position of second Mach disk and subsequent compression/expansion regions) in the upstream direction. In the plane corresponding to the new PIV measurements at $x/D=3.62$, the differences between predicted and measured average and rms velocity levels however remain typically less than $7\%$ and $10\%$ respectively. 

%\begin{figure}[htbp]
%\centering
%\includegraphics[width=2.8in, trim={0 0 10cm 4cm}, clip]{figures/PlanLong_Expe_DDES}
%\includegraphics[width=3in]{figures/PlanLong_Expe_DDES}
%\caption{Average streamwise velocity field: reference x-y %PIV data (surrounded by black rectangular box) versus overall numerical prediction. Position of the y-z cross-PIV plane (vertical black line). \textcolor{red}{CHANGE THE X/D value to 3.62}} 
%\label{figure-comparison-velocity}
%\end{figure}

\begin{figure}[htbp]
\centering
\includegraphics[width=3in]{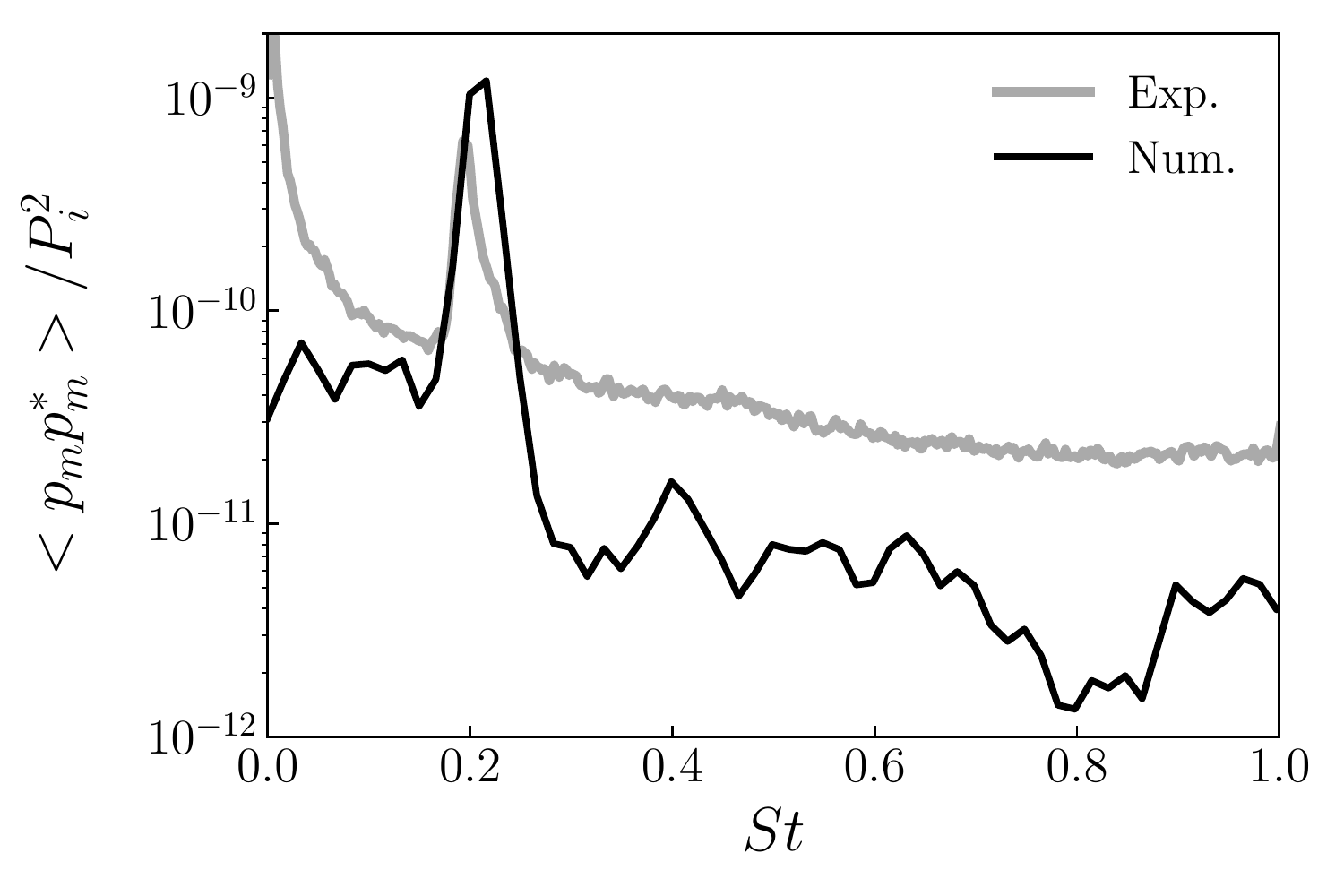}
\caption{\label{mode-expe-num} Comparison of the experimental and numerical PSD of second azimuthal pressure modes $m=1$ taken at $x/D=1.25$.}
\end{figure}

As detailed in following section \ref{unsteadiness-organization}, the fluctuating variables can be decomposed into azimuthal Fourier modes (see equation \ref{azimdecomp}).
The Power Spectral Densities (PSD) for the antisymmetric azimuthal Fourier mode $m=1$ of wall pressure fluctuations is presented in Fig.~\ref{mode-expe-num} for the position $x/D=1.25$ to compare with similar data available from the ring located at this position during the experiment.
The energy distribution of wall pressure fluctuations in the separation region is classically characterized by a large bump at low Strouhal number, mainly associated to the upstream motion of shock/separation line system, and contributions at high Strouhal number, associated to coherent structures advected within the jet mixing layer. Even if the upstream shock/separation line motion mainly signs on the first axisymmetric mode $m=0$, it has a non-negligible contribution to other modes, including this non-axisymmetric mode $m=1$. It is worth noting that the present evaluation of low-frequency content is expected to be less converged with simulation data (acquired during a more limited time duration) while a slight shift of upstream conditions during the experiment is naturally likely to exaggerate the very low-frequency content  experimentally evaluated. The under-estimation of high-frequency content of fluctuations during the simulation is also consistent with the observation of the delayed development of turbulence within the mixing layers and the lower resolution of the amplitude of pressure fluctuations radiated from these smallest scales of the flow.
However, the most dynamically active scales associated to flow oscillations in the intermediate frequency range appear to be particularly well reproduced. 
%These common features of PSD can be observed at any given streamwise position. The intensity of the various contributions naturally evolves as a function of the relative distance between the streamwise location considered and the separation line and as a function of the relative width of the mixing layer and its proximity to the nozzle wall. Low-frequency contributions are thus less pronounced for increasing distance in the downstream direction while the contrary is observed for contributions at higher frequencies associated to the turbulent structures developing within the mixing layer.
The very particular feature here observed is the presence of the tonal peak of the azimuthal mode  $m=1$ at $St \simeq 0.2$. A value of $St=0.216$ is predicted with the present simulation strategy, in rather good agreement with the experimental value of $St=0.198$. This indicates that the main global flow behavior responsible for the tonal behavior of most interest for the present study is well captured with the present simulation and that its relative importance for generation of lateral forces can be estimated.

%Although, the limited simulation time available leads to a less converged spectrum than the experimental one, the salient peak of mode $m=1$ in the intermediate frequency range, which is of main interest in the present study, appears well reproduced. These comparisons thus globally show that the present simulation strategy also enables to capture than main important dynamical features of flow.

As a conclusion, even if the present numerical strategy admittedly leads to a limited representativity of the simulated external flow entrainment and far-field jet dynamics, the large scales dynamical features of the jet appear to be satisfactorily reproduced in the whole separated region and close to the nozzle exit, allowing to conduct a finer analysis of the flow dynamics with confidence.

%%%%%%%%%%%%%%%%%%%%%%%%%%%%%%%%%%%%%%%%%%
%%%%%%%%%%%%%%%%%%%%%%%%%%%%%%%%%%%%%%%%%%
\section{Analysis of unsteadiness}
\label{unsteadiness}
\subsection{Global spatial and temporal organization of fluctuating field}
\label{unsteadiness-organization}

\begin{figure}[htbp]
\centering
\includegraphics[width=3in]{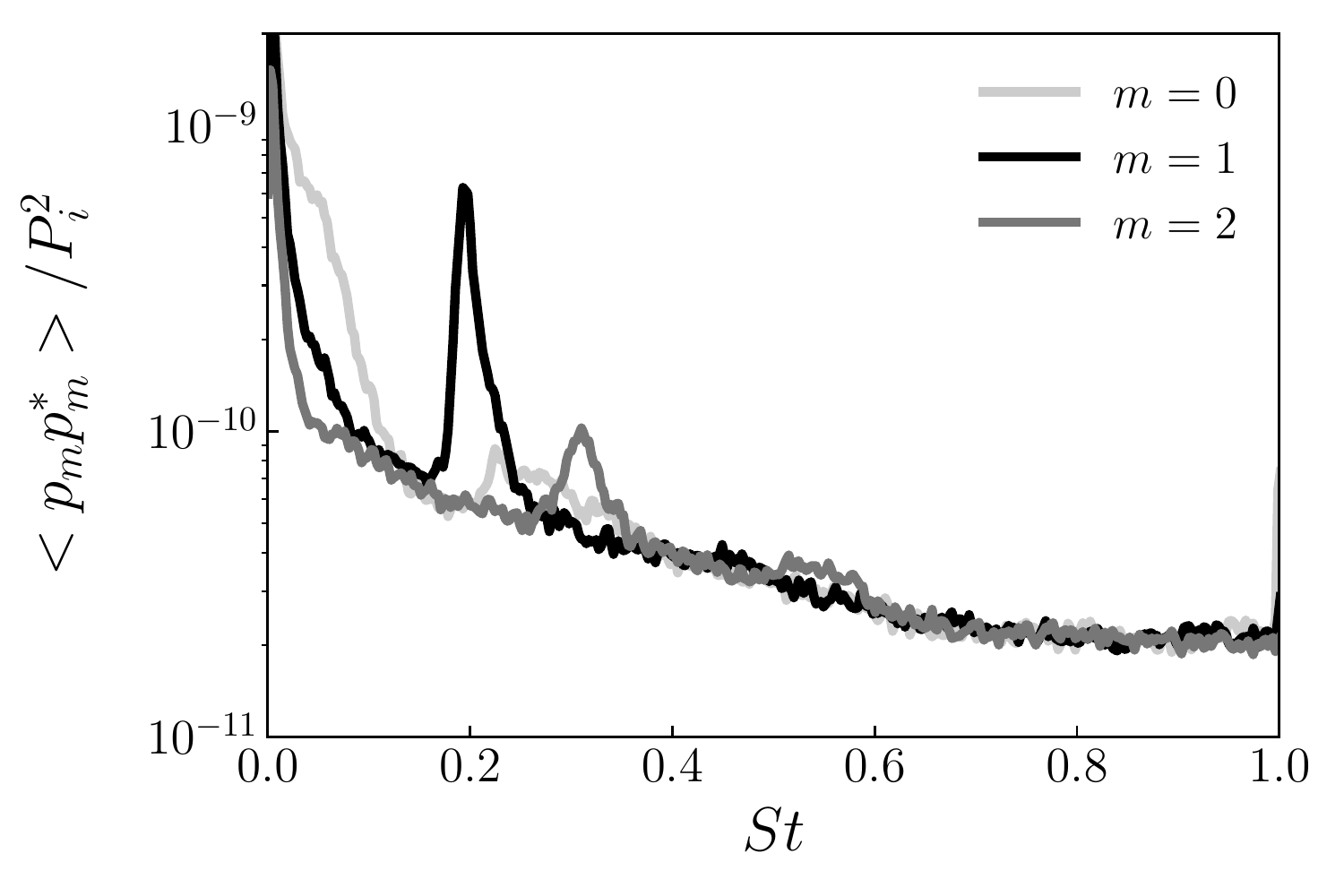}
\caption{\label{mode-expe} PSD of 3 first azimuthal pressure modes at $x/D=1.25$ obtained from the experiments.}
\end{figure}

In the following, we focus on the physical variables of most interest, including the pressure $p$, density $\rho$ and velocity components in streamwise $u_x$, radial $u_r$ and azimuthal $u_{\theta}$ direction. The fluctuating field is obtained by susbtracting the time averaged field to each instantaneous field. The spatial organization of this fluctuating field is then characterized by decomposing the  vector composed of a subset of any fluctuating physical variable $q=(p^{\prime},u_x^{\prime},u_r^{\prime},u_{\theta}^{\prime},\rho^{\prime})^T$ into azimuthal Fourier modes:
\begin{equation}
\label{azimdecomp}
q_m(x,r,t) = \frac{1}{2\pi}\int_0^{2\pi} q(x,r,\theta,t) e^{-i m \theta} d\theta,
\end{equation}
where $m$ is the azimuthal mode number.
The features of the fluctuating wall pressure field are first examined.
For each ring of wall pressure sensors, the pressure field is thus decomposed into azimuthal Fourier modes according to equation (\ref{azimdecomp}).
%\begin{equation}
%\label{eqazim}
%    p_m(x,t)=\frac{1}{2\pi} %\int_{0}^{2\pi} p(\theta,x,t)e^{-i m %\theta} \mathrm{d}\theta,
%\end{equation} 
%where $m$ is the azimuthal mode number. 
The PSD of the $3$ first modes evaluated at $x/D=1.25$ are plotted for example in Fig.~\ref{mode-expe}. The large low-frequency bump is visible on each distribution. This bump is all the more dominant as we consider a streamwise location close to the separation line and can be associated to the separation shock motion \cite{jaunet2017wall}. The large dominance of energy for $m=0$ close to the separation line suggests that this shock motion mainly remains axisymmetric. Some significant fluctuating energy is also contained in the high-frequency ($St > 0.8$) range and increases as we move in the downstream direction. It is attributed to the passage of smaller coherent structures advected along the separated region and jet mixing layer. Their contribution appear rather uniformly distributed in each azimuthal mode. The most salient feature is the presence of a peak of the antisymmetric mode $m=1$ at $St \simeq 0.2$ in the middle frequency range. It should be recalled that this peak exists for a narrow range of operating conditions but is more particularly dominant for $M_j=2.09$. Other peaks at other frequencies can be detected in the other azimuthal modes but remain of significantly lower amplitude.

\begin{figure}[htbp]
	\centering
	\includegraphics[width=3in]{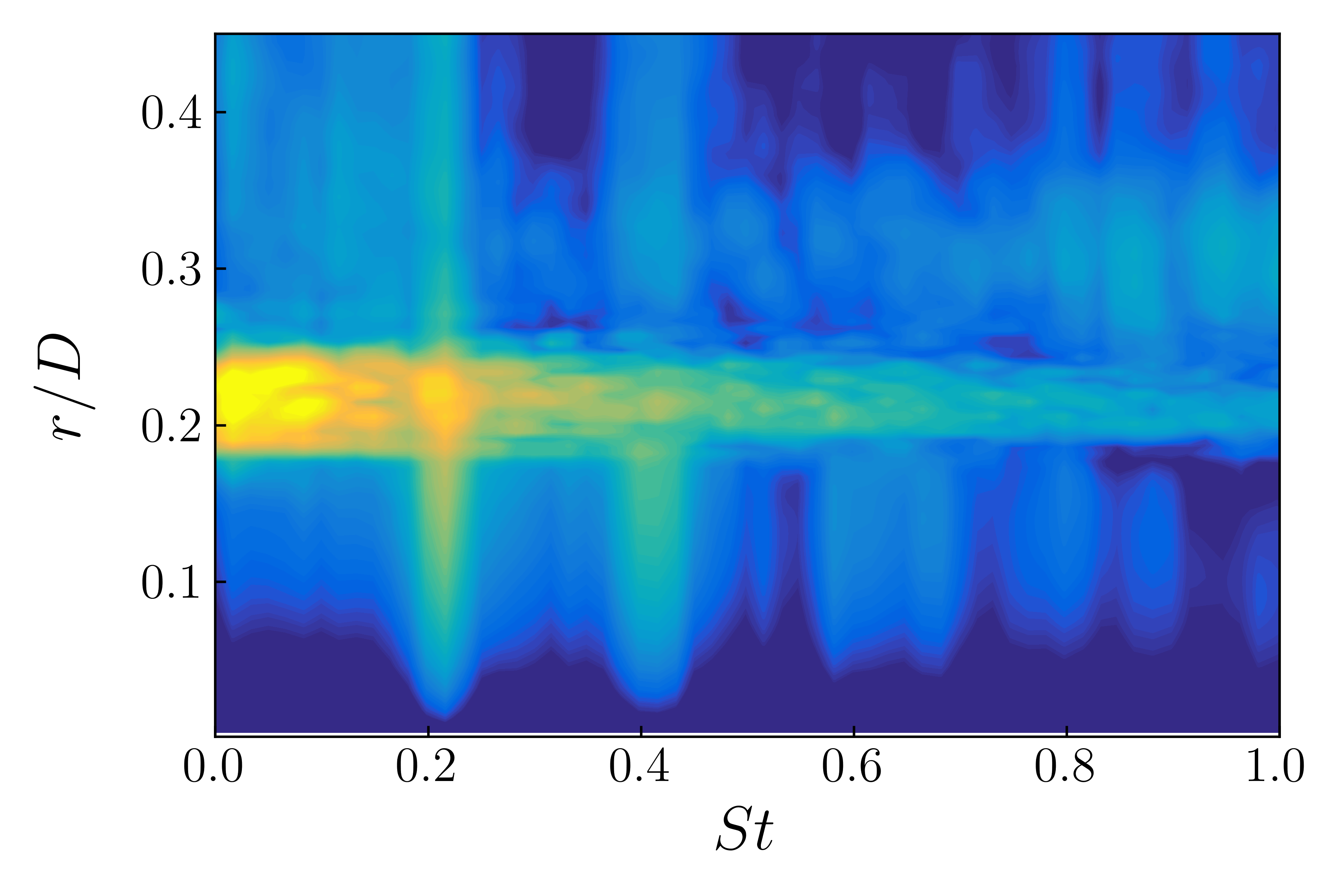}
	\caption{PSD map of the azimuthal pressure mode $m=1$ in the plane at $x/D=1.43$ as a function of the radius.}
	\label{figure12}
	\end{figure}

The present analysis is extended to the internal flow region based on numerical data. The PSD of the fluctuating pressure mode $m=1$ is for example shown in Fig.~\ref{figure12} for the plane $x/D=1.43$ (corresponding to the middle of the open separated region). As expected, some high levels of energy in a wide frequency range are observed between the radial positions $r/D=0.2$ and $0.25$. This zone corresponds to the mixing layer region spatially developing between the near axis subsonic jet core (downstream of the first internal Mach disk) and the external recirculation region. Whatever the radial position, a peak of energy also clearly emerges at $St\simeq 0.2$. 
This result thus indicates that the particular wall pressure field organization previously observed is not confined to the near-wall region. It rather appears as the signature of a more global flow organization of the pressure field prevailing through the whole separated flow region within the nozzle. The maximal energy levels are located within the mixing layer, more particularly close to the internal part of the mixing layer (at $r/D\simeq 0.215$) for this streamwise location. This highlights the probable prominent role of intrinsic non-axisymmetric instabilities spatially developing within the jet mixing layer inside the nozzle.

\begin{figure}[htbp]
	\centering
	\includegraphics[width=0.8\textwidth]{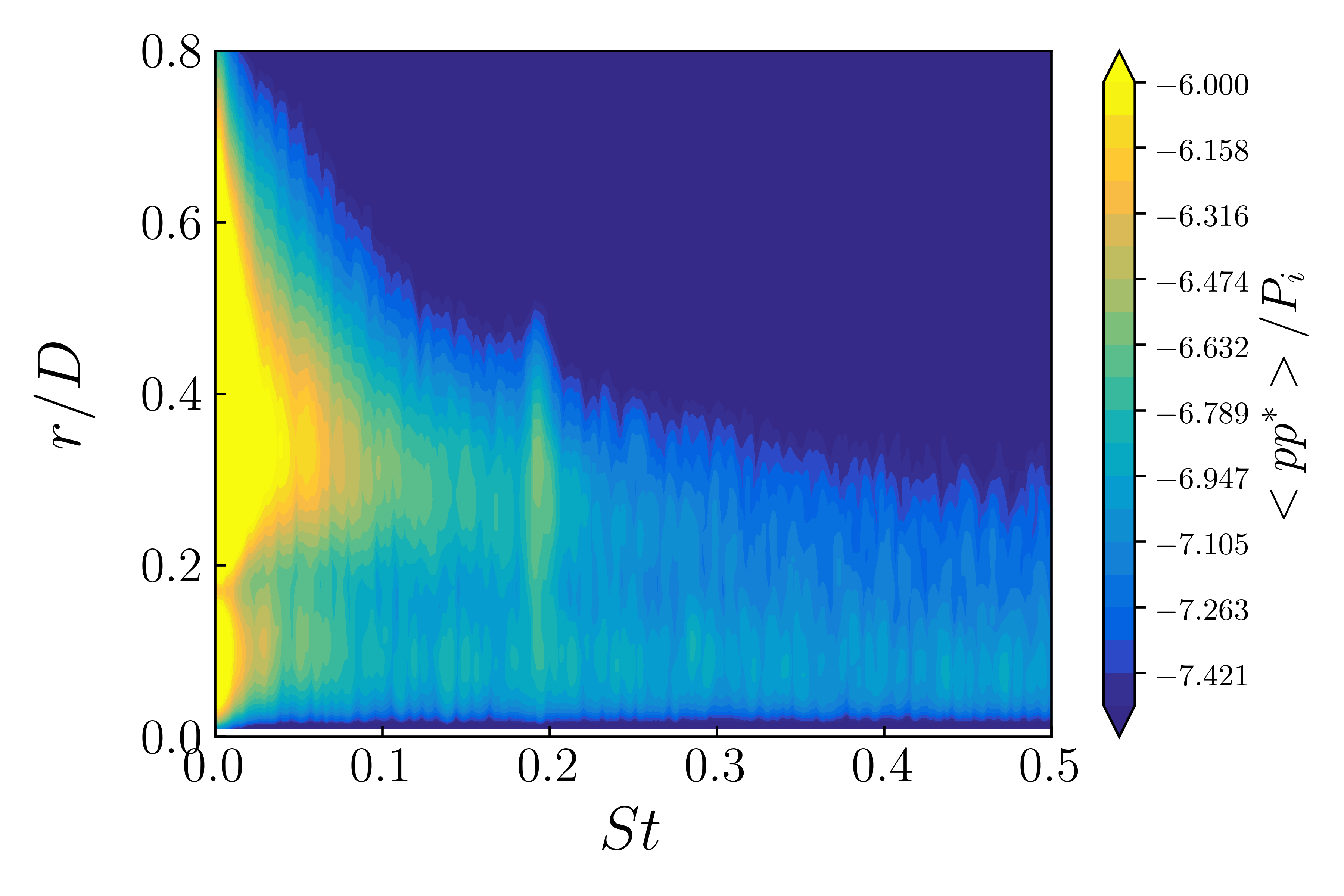}
	\caption{PSD map of the anti-symmetric azimuthal mode of velocity fluctuations (from PIV measurements), normalized by the fully expanded jet velocity, as function of radius and Strouhal number at $x/D=3.63$. Amplitudes are plotted using a logarithmic scale.}
	\label{figure-psd-velocity-plane}
	\end{figure}

The organization of the velocity field downstream of the nozzle exit and its link with the internal wall pressure field are now examined. The structure of velocity field at the position of the PIV plane (\textit{i.e.} $x/D = 3.63$) is first briefly described.
The overall flow visualization given by Fig.~\ref{figure-comparison-velocity} indicates that a second Mach disk forms just downstream of the nozzle exit and is followed by subsequent expansion/compression zones. The PIV plane is located close to the end of the pseudo-potential core slightly before the internal part of the mixing layer reaches the jet axis. 
In this region, a deficit of velocity could be still observed near the jet center in the radial profile of mean streamwise velocity. The mixing region roughly extends from $r/D=0.2$ to $0.6$. The flow at this location is composed of two distinct mixing regions: \textit{i/} an internal mixing layer issued from the triple point connected to the separation shock and first Mach disk within the nozzle and \textit{ii/} an external mixing layer between the surrounding supersonic coflow and the recirculation zone. Higher values of RMS velocity are naturally observed in this region with a peak value found near $r/D=0.33$, which corresponds to the merging zone between these internal and external mixing layers. %In order to analyze the azimuthal organization of the fluctuating velocity field in this plane, velocity data are interpolated on a cylindrical mesh with a resolution of $120$ points in the azimuthal direction before applying the Fourier spatio-temporal decomposition.  
As expected for such a position located quite far downstream, the results show that the fluctuating energy within the whole mixing region is distributed into a large number of azimuthal modes and within a large frequency range. However, as shown in Fig.~\ref{figure-psd-velocity-plane}, a small peak still clearly emerges at $St\simeq 0.2$ in the PSD of first azimuthal mode of velocity. It is more particularly visible near $r/D=0.33$ where the merging between the internal and external mixing layers is observed and where the maximal rms values are detected. The amplitude of this energy peak of first mode of velocity field is here far lower than the one of first azimuthal pressure mode of the internal wall pressure signals. It remains however sufficiently significant to suggest that the turbulent structures spatially developing within the jet mixing layers also exhibit the same signature of a common global oscillation mechanism.

\begin{figure}[htbp]
	\centering
	\includegraphics[width=3in]{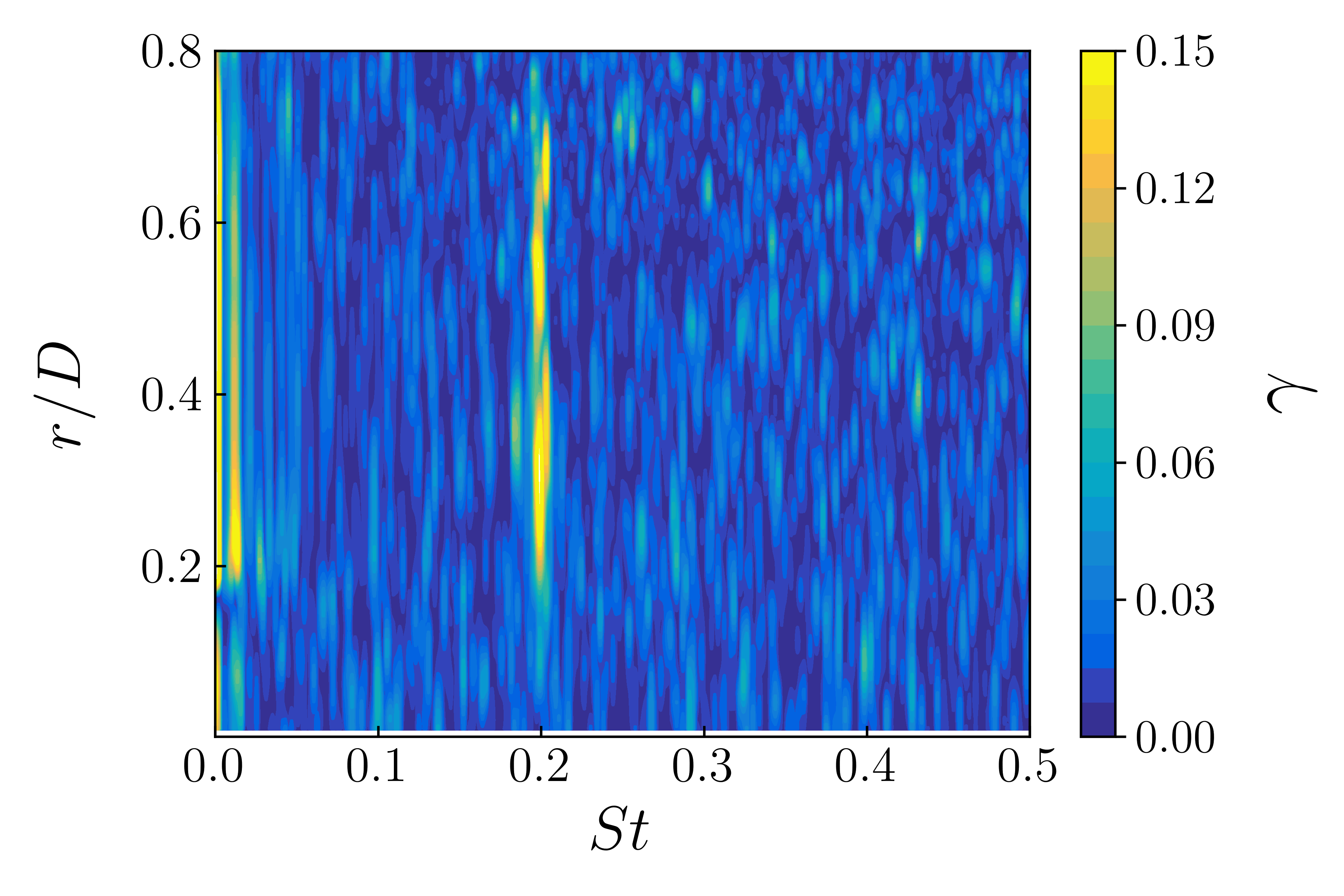}
	\caption{Coherence map between azimuthal modes $m=1$ of wall pressure signals at $x/D=0.9$ and jet velocity at $x/D=3.62$.}
	\label{figure-coherence}
	\end{figure}

Synchronized and time-resolved data of wall pressure and external velocity have been obtained during a time period long enough to allow an extended correlation analysis based on experimental measurements. A low-pass filtering of wall pressure data is first applied in order to obtain pressure and velocity signals with similar sample rate. The coherence between the first azimuthal modes of internal pressure at a ring of wall pressure sensors and first azimuthal mode of streamwise velocity is computed. The coherence map obtained remains rather similar whatever the pressure ring position considered. The map obtained for the ring of sensors at $x/D=0.9$ and the PIV plane at $x/D=3.63$ is for example presented in Fig.~\ref{figure-coherence}. As expected, the signals are uncorrelated at nearly all frequencies due to the important distance between the internal sensors and the external velocity plane. However, a significant coherence level emerges at $St \simeq 0.2$ and more particularly close to $r/D=0.33$ in the external velocity plane and to a lower extent close to $r/D=0.51$. As previously mentioned, this first position corresponds to the zone where the internal and external mixing layers start merging. The second outer position is close to the boundary of the external mixing layer. 
In spite of high-amplitude and stochastic turbulent fluctuations found within the jet at such a quite far downstream position, the existence of such a peak of coherence reveals that first modes of internal pressure fluctuations and external velocity fluctuations remain significantly correlated but only in a narrow middle frequency range. The non-axisymmetric modes of internal pressure field and the external jet velocity field thus share the common signature of an organized coherent motion in this particular narrow frequency range.

%%%%%%%%%%%%%%%%%%%%%%%%%%%%%%%%%%%%%%%%%%%%%%%%%%%%%%%%%%%%%%%
\section{Identification of global coherent jet flow structure}
\label{global-coherent-structure}

\subsection{Coherent structures eduction}

In order to explore the coherent structures of the flow, we propose to decompose the flow in orthogonal modes using Spectral Proper Orthogonal Decomposition (SPOD). Details of the SPOD have been recently discussed in \citep{towne2018spectral} and we provide in the following a brief recall
of the processing to introduce the notations.\\
In order to compute the SPOD modes $\Psi$, we first form the cross-spectral density (CSD) matrix $\mathbf{P}_{qq} \,$, at the frequency $\omega$,  of vector $q$ whose components correspond to a chosen subset of fluctuating parts of physical variables of interest. The dominant SPOD modes identified can sometimes differ as a function of the components chosen to build $q$. In the present case, it has been checked that considering one single component vector based on the fluctuating pressure or a vector composed of both the pressure fluctuation and fluctuation of the streamwise velocity component  leads to similar results. 
The results shown in the following are obtained based on the two components vector $q=(p^{\prime},u_x^{\prime})^T$.
$\mathbf{P_{qq}}$ is computed using Welch's periodogram method :
\begin{eqnarray}
\mathbf{P}_{qq} &= & E\left\{ \hat{q} \hat{q}^H\right\},
\end{eqnarray}
where $E$ denotes the expectation operator, $\hat{q}$ is amplitude of the time domain Fourier transform of $q$ at the frequency $\omega$:
\begin{eqnarray}
\hat{q} & = & \sum_t q(t) e^{-i \omega t},
\end{eqnarray}
and $\hat{q}^H$ is its transpose conjugate. Note that for reasons of clarity, we dropped the time and space dependence of $q = q(\mathbf{x},t)$ whenever possible.\\
Finally, the SPOD modes are obtained by computing the eigenvectors of the CSD matrix:
\begin{eqnarray}
  \mathbf{P}_{qq} \mathbf{\Psi} & = & \mathbf{\Psi} \mathbf{\Lambda},
\end{eqnarray}
where $\mathbf{\Lambda} = diag(\lambda^{(i)})$ is the diagonal matrix of the SPOD eigenvalues sorted in ascending order. Since $\mathbf{P}_{qq}$ is Hermitian by construction all eigenvalues are positive. $\mathbf{\Psi}$ is the matrix whose columns are the individual SPOD modes $\Psi^{(i)}$:
\begin{eqnarray}
  \mathbf{\Psi} & = & \left[ \Psi^{(1)}, \Psi^{(2)}, ... \, , \Psi^{(N)} \right],
\end{eqnarray}
where $N$ is the rank of $\mathbf{P_{qq}}$.\\
A low order reconstruction of the CSD matrix $\tilde{\mathbf{P}}_{qq}$ can be formed using a subset of $N_{tr}$ SPOD modes and eigenvalues:
\begin{eqnarray}
\tilde{\mathbf{P}}_{qq} &= & \sum_{i=1}^{N_{tr}} \Psi^{(i)} \lambda^{(i)} {\Psi^{(i)H}},
\end{eqnarray}
In the following, we focus on the first non-axisymmetric azimuthal mode of the flow fluctuations $q_m(x,r,t)$ with $m=1$, since this is the one supporting most of the dynamics at the frequency of the tonal dynamics. The flow variables are therefore first decomposed into azimuthal Fourier modes, as given by eq. (\ref{azimdecomp}), prior to compute the SPOD modes,
and we retain only the $m=1$ mode $q_1(x,r,t)$. Note that, for sake of conciseness, we dropped the $(\cdot)_1$ under-script in the remaining of the paper.

\begin{figure}[htbp]
	\centering
	    %\subfloat[St=0.060]{
	    \includegraphics[width=0.7\textwidth]{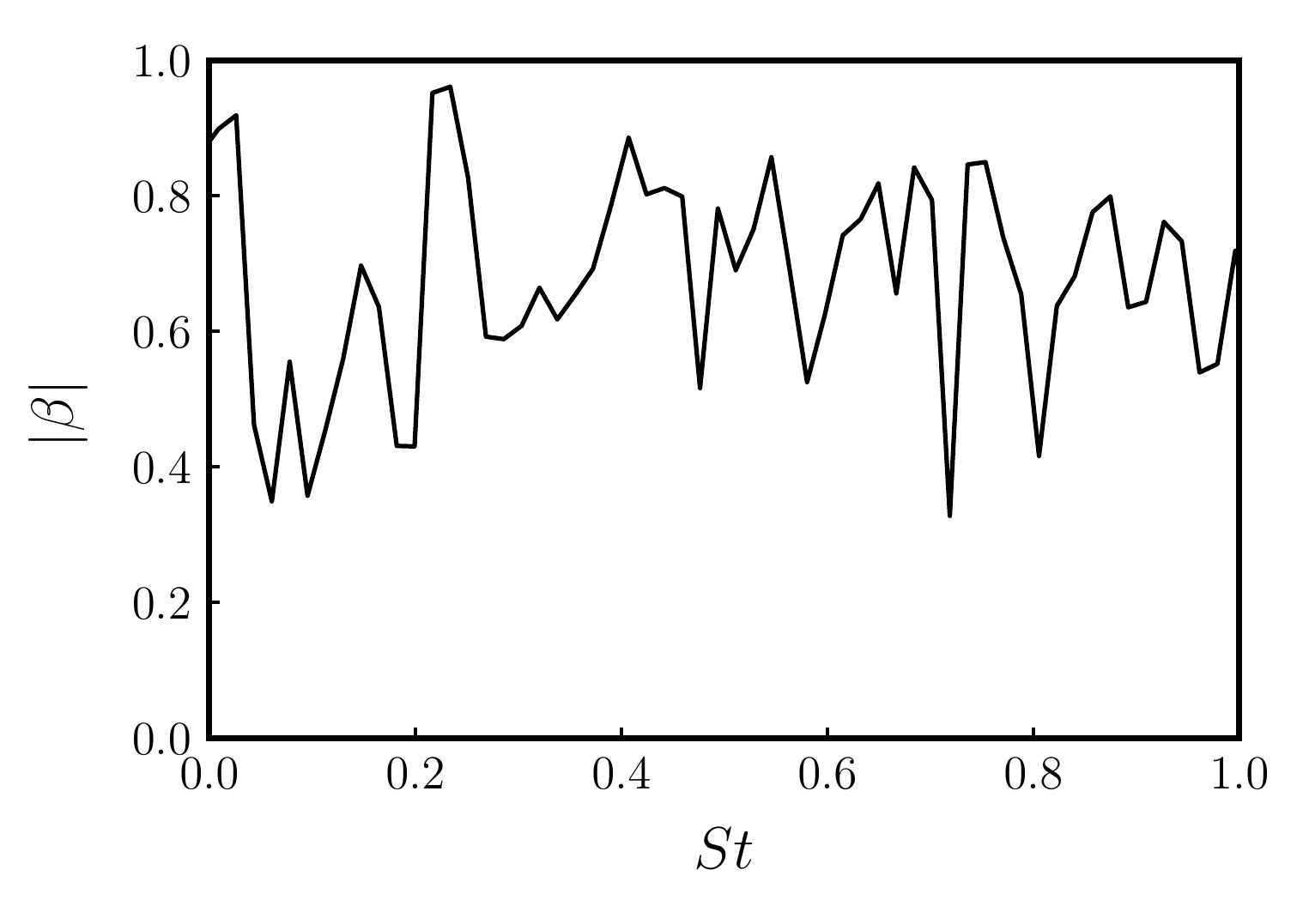}
	    %}\\
	    	\caption{Absolute value of the scalar product $\beta$ between the first SPOD modes computed with the first half of the data and the ones computed with the second half of the data.}
	\label{SPOD:convergence}
	\end{figure}

Since they are based on a two-points statistics of the flow, the computation of SPOD modes requires a large amount of data. In order to control the convergence rate of the decomposition and ensure that the analysis presented in the following sections is relevant, we separate the data in two halves and SPOD modes are computed on each subset. Then, we compute the scalar-product $\beta$ between the SPOD modes obtained with each of these subsets:
\begin{eqnarray}
\beta = (\psi^{(i)}_1,\psi^{(i)}_2),
\end{eqnarray}
where $(\cdot,\cdot)$ is the scalar product, $\psi^{(i)}_1$ and $\psi^{(i)}_2$ refer to the SPOD modes obtained with the first and the second half of the data respectively.\\ 
For a pair of SPOD modes at one given frequency, a scalar product differing from unity reveals the non-similarity between each corresponding mode of each database. The analysis of the spatial support of the mode is likely to be biased in such a case. In the present study, it is indeed found that the original set of snapshots is not large enough to ensure a convergence of the second and higher order SPOD modes. Increasing the size of the database with longer simulation times (which was not possible due to computational limits) would be required to improve the convergence rate. The convergence of first SPOD mode however seems to be sufficient at the tonal frequency. The scalar product between this first SPOD modes extracted from each halves of the database is plotted against frequency for the available frequency range in Fig.~\ref{SPOD:convergence}. As can be seen, the average scalar product is around $\beta=0.6$, indicating that even the convergence of the first SPOD mode is not reached for many frequencies. On the contrary, a value very close to unity is measured at the tonal frequency of interest of $St \simeq 0.2$. Hence, the SPOD mode at this frequency can be reliably examined. The relative difference of energy content of the other higher modes with respect to the energy of first mode can still be reliably interpreted but their too weak convergence rate \textit{a priori} inhibits a clear interpretation of their spatial shape which would be likely to evolve by considering more snapshots for the analysis. Only the first mode at the tonal frequency will thus be considered into detail in the following analysis and the spatial shape of higher modes will not be discussed.

\subsection{Dominant SPOD mode}
\begin{figure}[htbp]
	\centering
    \includegraphics[width=0.7\textwidth]{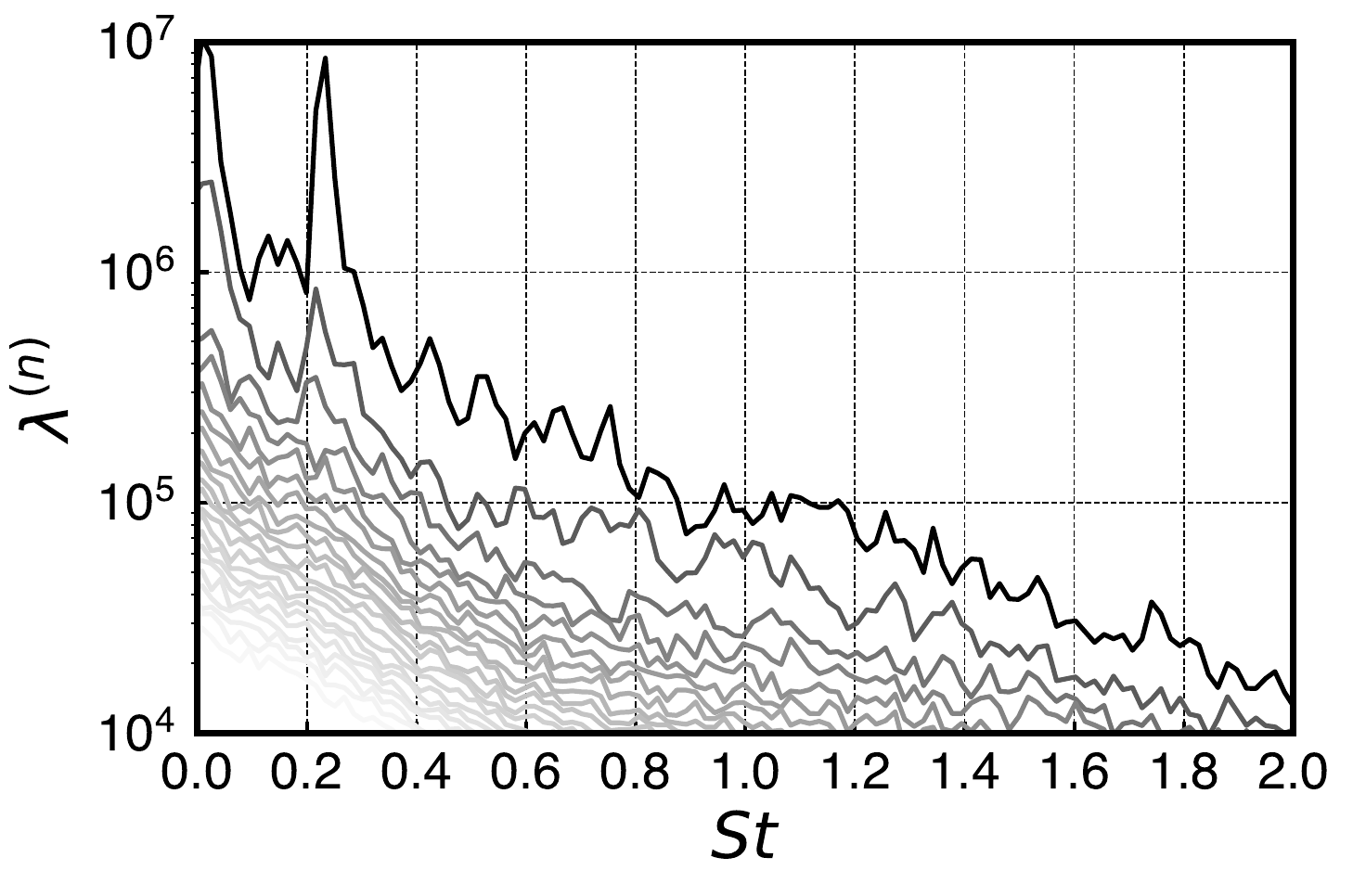}
   	\caption{SPOD eigenvalue as function of Strouhal number, darker lines correspond to the more energetic eigenvalues.}
	\label{SPOD:eigenvalues}
\end{figure}

We present in Fig.~\ref{SPOD:eigenvalues} the eigenvalues of the SPOD modes as function of the Strouhal number. The energy content of the first mode clearly dominates the energy content of all the other modes at any frequency (with a ratio varying typically between $2$ to $4$). It is striking that the resonant frequency SPOD modes show a far more important gain separation (of more than one decade) between the first and second SPOD modes. This emphasizes that the tonal dynamics differs in nature from the rest of the spectrum. At the resonant frequency, the first SPOD mode represents more than 80\% of the energy of the flow dynamics. This shows that most, if not all, of the dynamics associated with the resonance has been extracted from the data by the SPOD analysis. 

%\begin{figure}[htbp]
%	\centering
%    \includegraphics[width=0.7\textwidth]{figures/spodeigvalues_p_ux_percent.eps}
%   	\caption{Fraction of the first SPOD mode energy with respect to the overall energy.}
%	\label{SPOD:fraction}
%\end{figure}    

\begin{figure}[htbp]
	\centering
	    %% \subfloat[St=0.22]{
	        \begin{tabular}[b]{c}%
	        \includegraphics[width=0.9\textwidth]{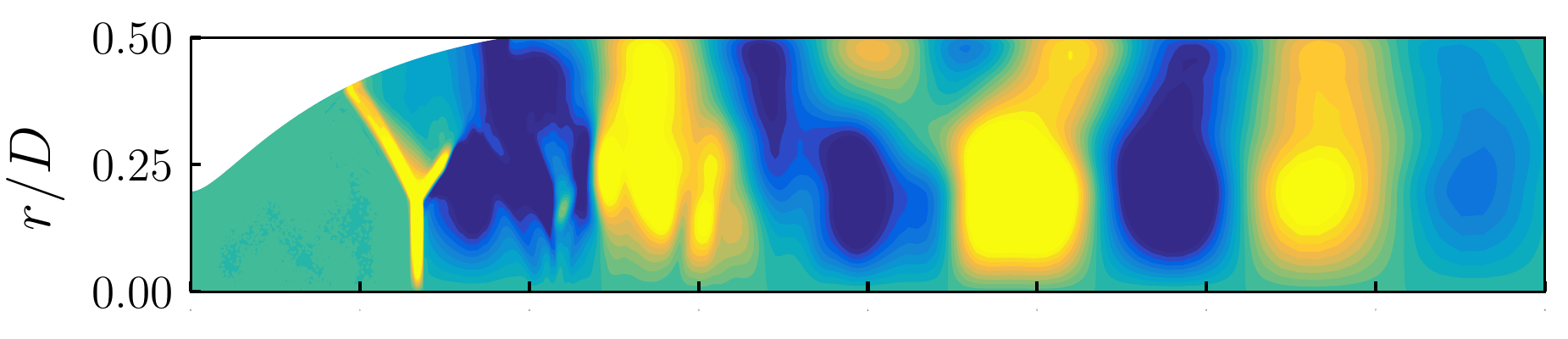}\\
	        \includegraphics[width=0.9\textwidth]{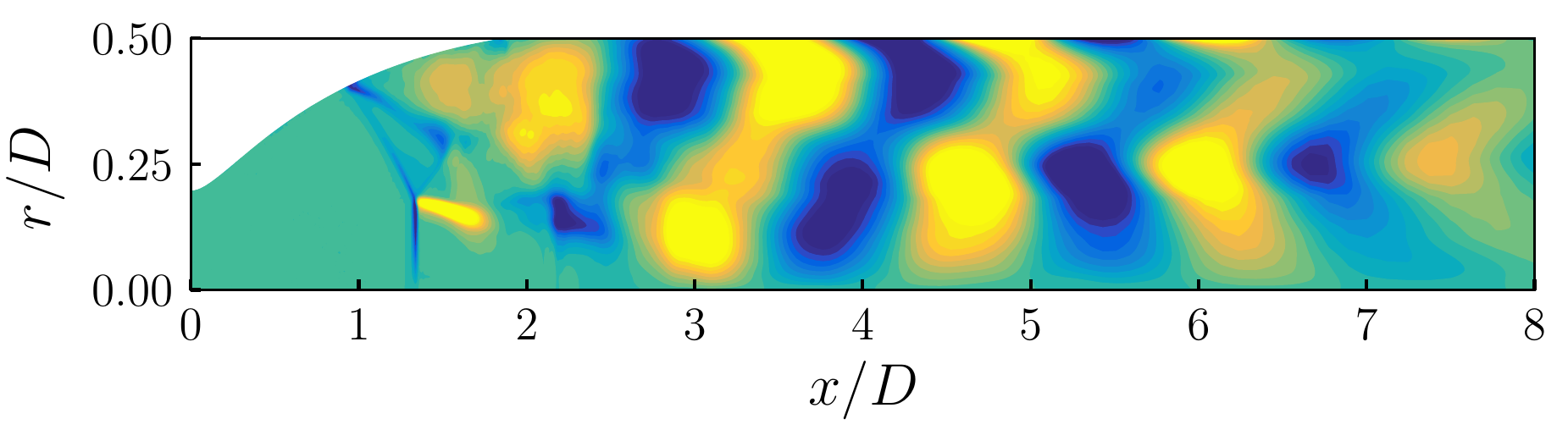}
	        \end{tabular}
	    %%}\\
	\caption{Real part of the first SPOD mode at $St = 0.216$: pressure (top) and streamwise velocity (bottom). The amplitude have been normalized by the maximum value and the color scale is saturated in $\pm 20 \%$ of this maximum value.}
	\label{SPOD:spodmodes_global}
\end{figure}

The dominant SPOD mode spatial signature at the Strouhal number of the resonance is presented in Fig.~\ref{SPOD:spodmodes_global}. There are several important structures to notice in the spatial support of this SPOD mode. First of all, we recognise inside the nozzle the fingerprint of the adaptation shock structure with its Mach disk within the nozzle. This shows that the shock system motion naturally shares the signature of the dynamics at the resonant frequency. Note that such discontinuities are expected to be excessively highlighted since they impose amplitude of local pressure and velocity gradients far more important than the one associated with other travelling coherent structures in the flow. Note that the color palette has been thresholded so as to avoid that these shocks hide the presence of other essential elements of the global structure.

Also visible in the figure is the presence of two distinct large scale wavepackets. A first inner wavepacket appears partially confined inside the exhaust jet, \textit{i.e.} for $ r/D < 0.25 $ approximately. The second outer wavy pattern is also recognizable beyond the jet boundary for $ r/D > 0.25 $. The presence of these wavy patterns in the SPOD mode structure all along the jet confirms the link previously established between internal and external fluctuations. In order to better identify the possible origin of the tonal behavior, the propagative properties carried by this SPOD mode are examined in the following section.

%$L_{shock}= 0.806 D$\\
%$k^- = -0.666 D$\\
%$k^+ = 0.8325 D$\\

\subsection{Upstream/downstream propagating waves}

Animating this dominant SPOD mode shows that the inner wavepacket possesses wave fronts moving in the downstream direction whereas the outer wavepacket is propagating in the upstream direction. Note that we cannot infer on the group velocity of such structures and cannot put directly into evidence that the energy carried by those waves goes in the same directions. 
The observation of the phase velocity directions of these two wavepackets is however already by itself a strong indication of a possible emergence of resonance.

\begin{figure}[htbp]
	\centering
	    %% \subfloat[St=0.22]{
	        \begin{tabular}[b]{c}%
	        \includegraphics[width=0.9\textwidth]{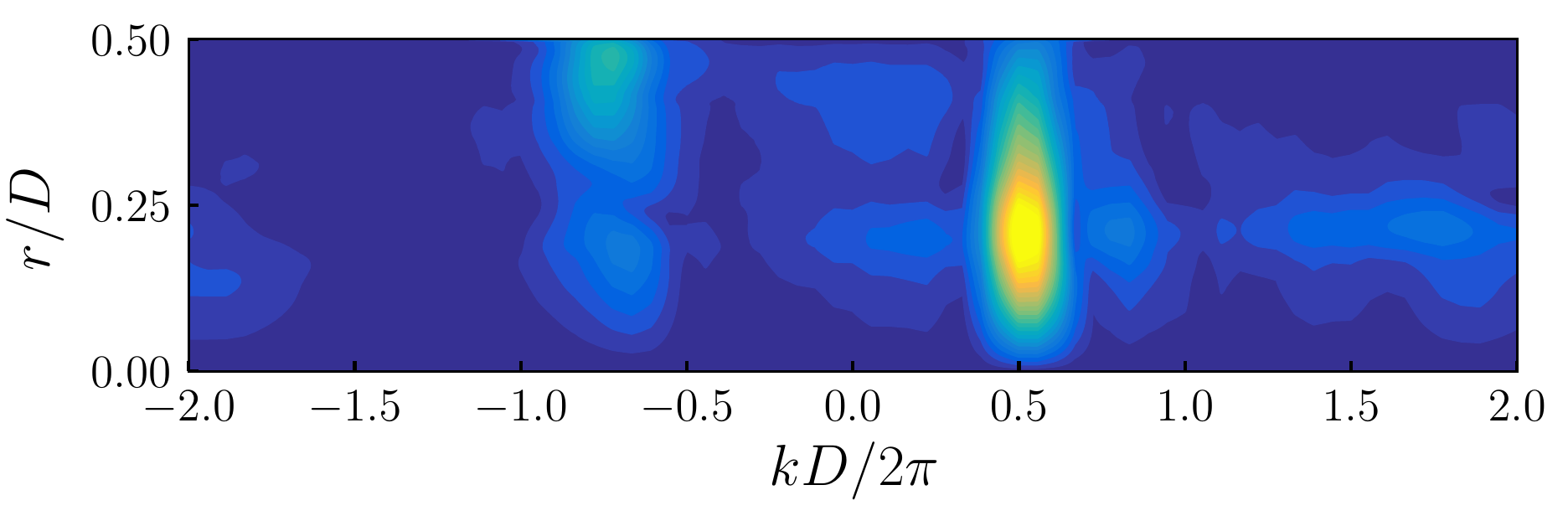}\\
	        \end{tabular}
	    %%}\\
	\caption{Fourier transform SPOD mode of pressure at $St = 0.216$. The amplitude have been normalized by the maximum value and the color scale is saturated in $\pm 20 \%$ of this maximum value.}
	\label{SPOD:spodmodes_fourier_k}
\end{figure}

Following the analysis proposed by \citet{edgington2018upstream}, the essential ingredients composing the dominant SPOD mode at the resonant frequency are further examined by carrying out a Fourier transformation in the streamwise direction:
\begin{equation}
    \hat{\Psi}^{(1)}(k,r) = \sum_x \Psi^{(1)}(x,r)e^{ikx},
\end{equation}
where $k$ is wavenumber in the streamwise direction. For each radial position, the window considered for the Fourier transform extends from the first available point at the nozzle wall up to the end of the computational domain. We present in Fig.~\ref{SPOD:spodmodes_fourier_k} the amplitude distribution of $\hat{\Psi}^{(1)}(k,r)$. It is evident, from this figure, that the SPOD modes spectrum is dominated by two distinct regions: one in the positive wavenumber domain ($k \simeq 0.5$), \textit{i.e.} with positive phase speed, and the other in the negative wavenumber region ($ k \simeq -0.7$),  \textit{i.e.} with negative phase speed. This confirms the observations we made while animating the SPOD mode in the time domain.

\begin{figure}[htbp]
	\centering
	    %% \subfloat[St=0.22]{
	        \begin{tabular}[b]{c}%
	        \includegraphics[width=0.9\textwidth]{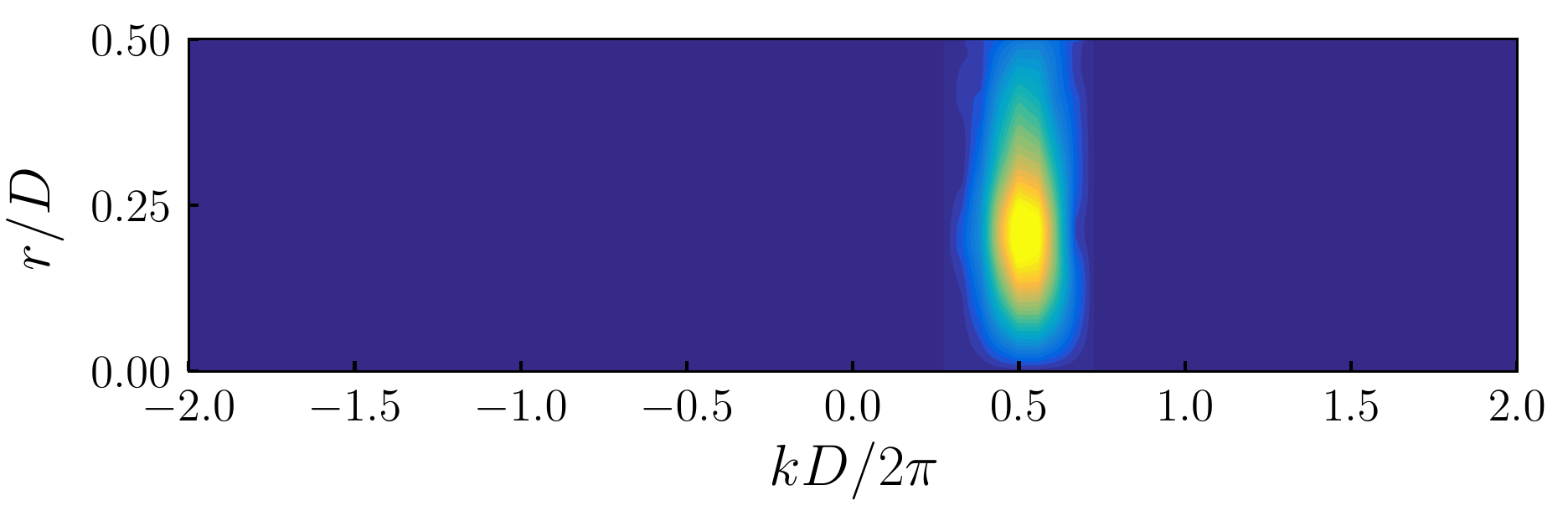}\\
	        \includegraphics[width=0.9\textwidth]{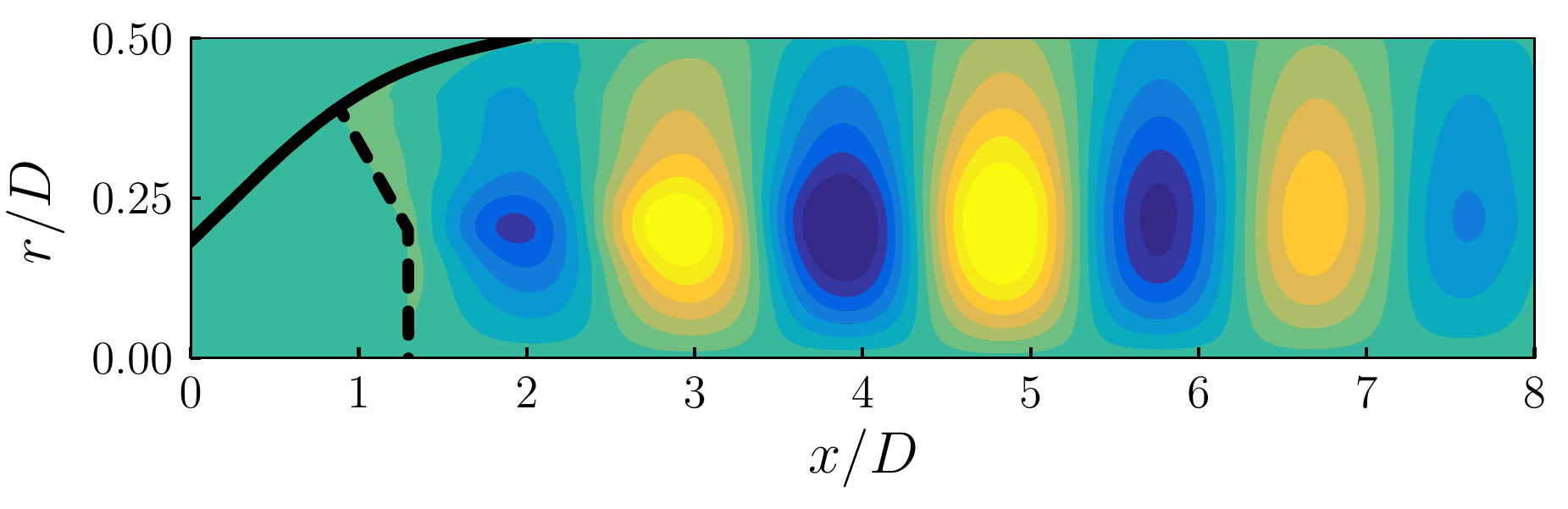}\\
	        \end{tabular}
	    %%}\\
	\caption{Positive phase-speed filtered Fourier transform (top) and its inverse (bottom), of the first SPOD mode of pressure at $St = 0.216$. The amplitude have been normalized by the maximum value of the original, non-transformed, SPOD mode and the color scale is saturated in $\pm 20 \%$ of this maximum value.}
	\label{SPOD:spodmodes_filtered_recons_pos}
\end{figure}

In Fig.~\ref{SPOD:spodmodes_filtered_recons_pos} (top) we present a filtered version of the previous Fourier spectrum where we have only retained the wavenumbers around the dominating downstream traveling waves. The corresponding pressure waves in the physical domain are plotted in \ref{SPOD:spodmodes_filtered_recons_pos} (bottom). The streamwise organisation of this structure, \textit{i.e.} growth, saturation and decay, as well as its radial shape, \textit{i.e.} exponential radial decay, is strong indication that this structure is similar to a Kelvin-Helmholtz instability wave. Interestingly, the peak amplitude of this structure is located at the level of the internal mixing layer issued from the triple point downstream of the shock structure, suggesting that this wave is probably generated by a classical inflectional mechanism, but here located in this internal part of the jet. The external part of the shear layer (between the the separation line and the impinging point of the reflected shock) does not participate to the wavepackets here identified. These observations thus put forwards the role of the internal mixing layer in the resonance mechanism of this over-expanded jet and suggest that the sensitivity zone for the emergence of the resonance is located near the triple point rather than the separation line as suggested in recent studies \cite{martelli2020flowdynamics}. In the rest of the paper, we will use the term KH wave when we will refer to this downstream traveling structure. 

\begin{figure}[htbp]
	\centering
	    %% \subfloat[St=0.22]{
	        \begin{tabular}[b]{c}%
	        \includegraphics[width=0.9\textwidth]{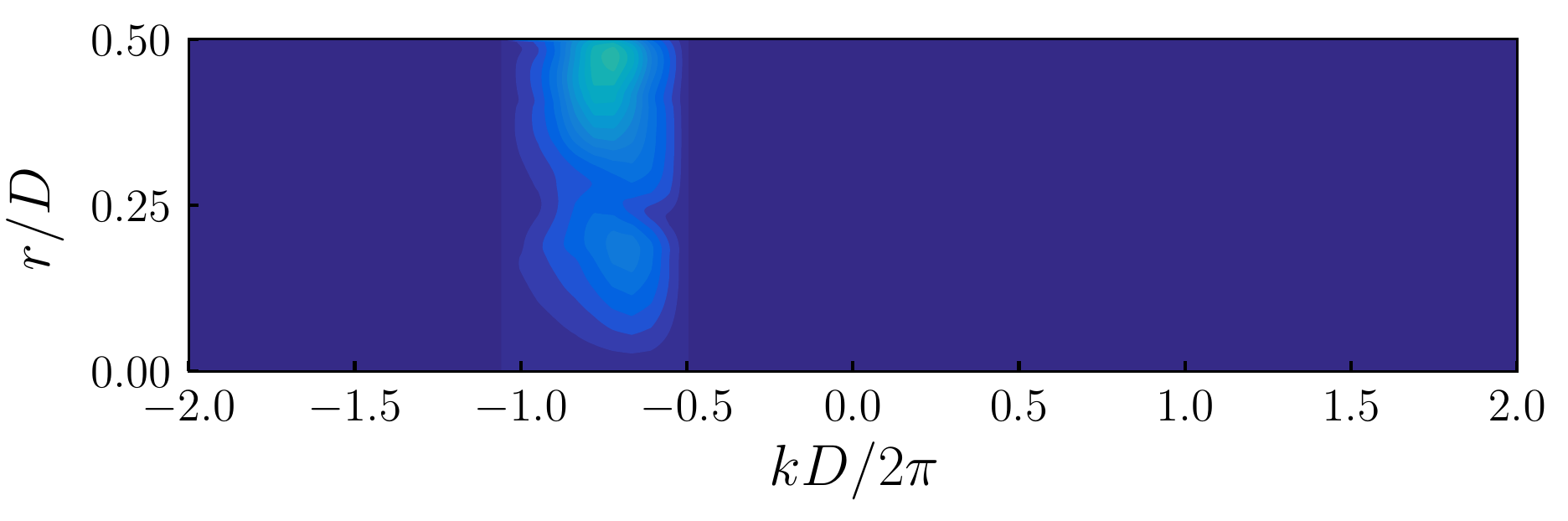}\\
	        \includegraphics[width=0.9\textwidth]{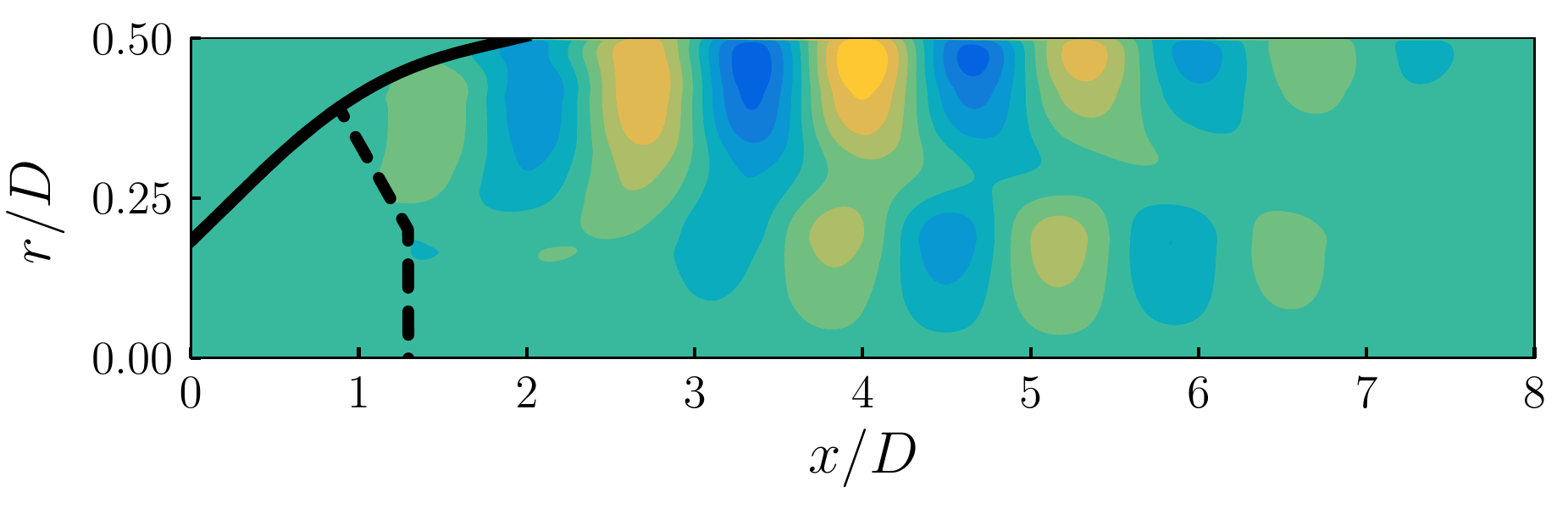}\\
	        \end{tabular}
	    %%}\\
	\caption{Negative phase-speed filtered Fourier transform (top) and its inverse (bottom), of the first SPOD mode of pressure at $St = 0.216$. The amplitude have been normalized by the maximum value of the original, non-transformed, SPOD mode and the color scale is saturated in $\pm 20 \%$ of this maximum value.}
	\label{SPOD:spodmodes_filtered_recons_neg}
\end{figure}

Similarly, in Fig.~\ref{SPOD:spodmodes_filtered_recons_neg} (top) we present a filtered version of the previous Fourier spectrum where we have only retained the wavenumbers around the dominating upstream traveling waves. The corresponding pressure waves in the physical domain is plotted in Fig.~\ref{SPOD:spodmodes_filtered_recons_neg} (bottom). It is evident that most of its energy is located outside of the jet, which makes it an excellent candidate to close the feedback loop of the resonance occurring at this frequency. An important feature to note, is that this coherent structure has some non-negligible signature inside the jet and more particularly within the internal mixing layer. Hence, it is possible that the downstream KH structure and this upstream wave exchange energy near by the adaptation shock triple point, where the KH receptivity is \textit{a priori} maximum. 

\begin{figure}[htbp]
	\centering
	    %% \subfloat[St=0.22]{
	        \begin{tabular}[b]{c}%
	        \includegraphics[width=0.9\textwidth]{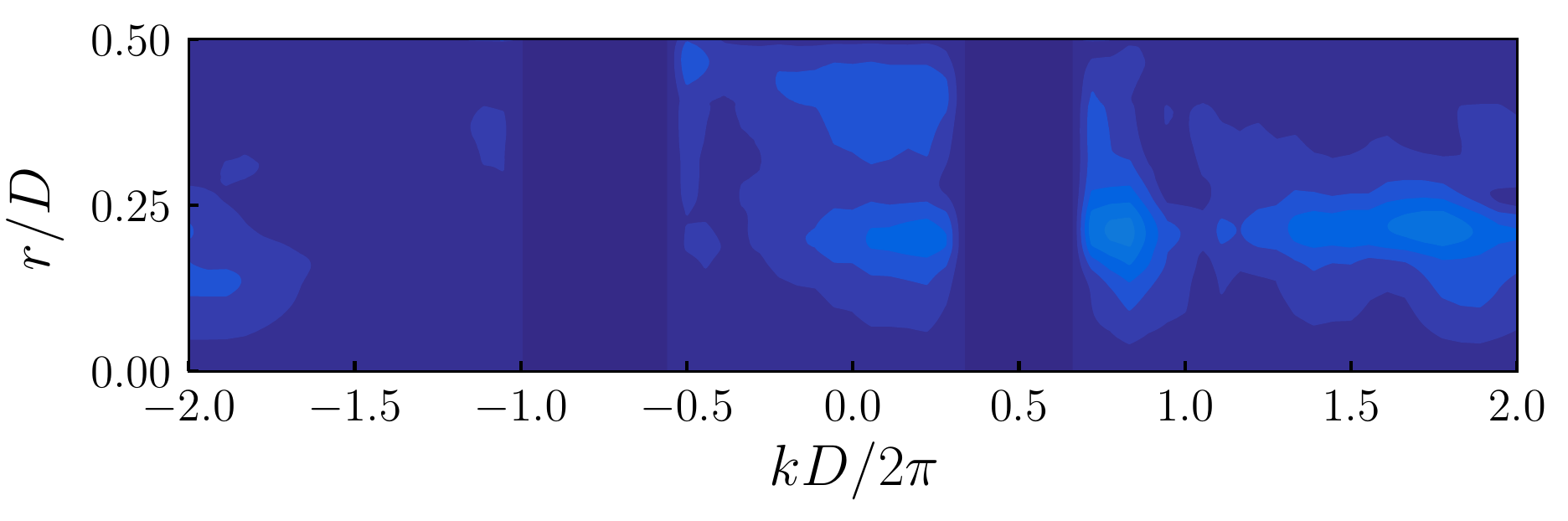}\\
	        \includegraphics[width=0.9\textwidth]{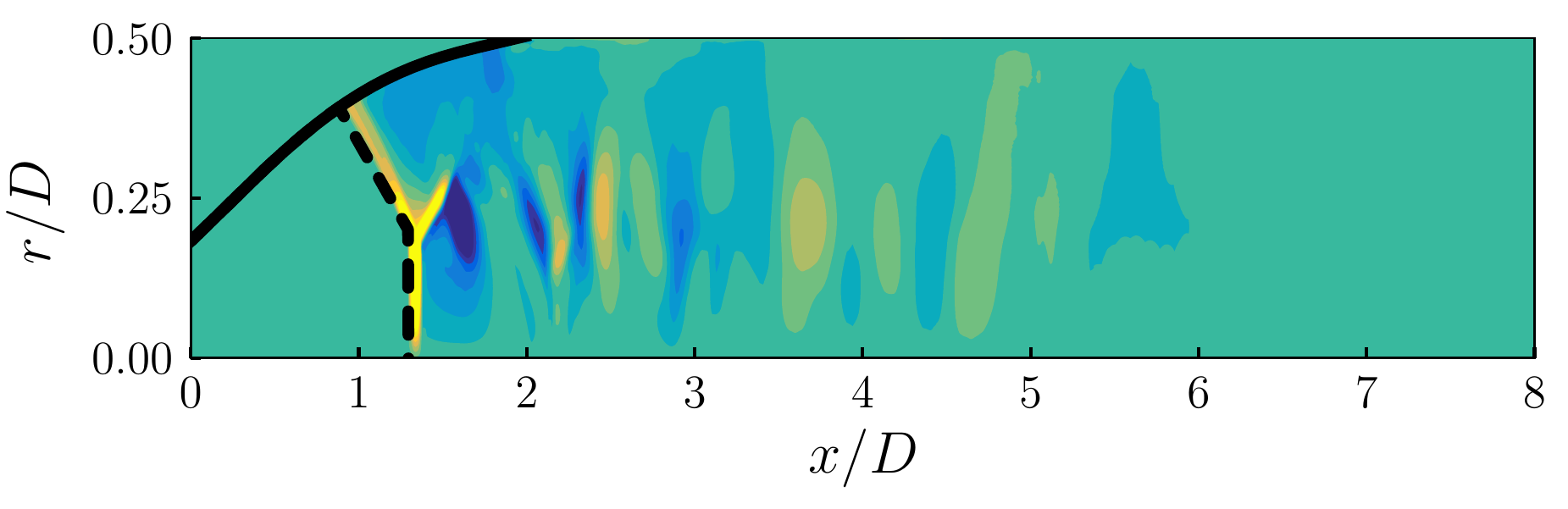}\\
	        \end{tabular}
	    %%}\\
	\caption{Remaining filtered Fourier transform (top) and its inverse (bottom), of the first SPOD mode of pressure at $St = 0.216$. The amplitude have been normalized by the maximum value of the original, non-transformed, SPOD mode and the color scale is saturated in $\pm 20 \%$ of this maximum value.}
	\label{SPOD:spodmodes_filtered_recons_shock}
\end{figure}

Finally, in Fig.~\ref{SPOD:spodmodes_filtered_recons_shock} (top) we present a filtered version of the previous Fourier spectrum where we have only retained the remaining wavenumbers after substraction of both the dominating upstream and downstream traveling waves. The corresponding pressure waves in the physical domain are plotted in Fig.~\ref{SPOD:spodmodes_filtered_recons_shock} (bottom). It is clear from the figure that the remaining wavenumbers represent the fluctuations associated to the motion of the shock structures (first and second Mach disks) as well as the train of other compression and expansion fans in the supersonic annular mixing layer of the jet column. Although these wavenumbers do not dominate the spatial Fourier spectrum, it shows, as expected, that the shock system responds at the resonance frequency in a very coherent manner.

The later observations allow to propose a plausible mechanism of the observed resonance. The most likely scenario involves a downstream unstable Kelvin-Helmholtz instability wave supported by the inner shear layer and an external wave, possessing support inside the core of the jet, to close the feedback loop. This underpins the conjecture given in \citet{jaunet2017wall} about the role of jet dynamics, and the internal shear layer, on the resonance. It makes, however, some important differences with the classical screech like scenario, where the outer shear layer supports both the downstream- and upstream-propagating disturbances (see for example \cite{tam1986proposed,gojon2018oscillation,edgington2018upstream,mancinelli2019screech}). Then, the resonance synchronizes all the jet dynamics, especially the outer shear layer, providing strong enough pressure fluctuations to make the shock system move to adapt to the pressure changes. Since the $m=1$ mode has both real and imaginary part by construction, it also possesses a negative frequency content which may differ from its positive counterpart. It has been checked that the same exact conclusions could have been drawn by studying the SPOD mode for the corresonding negative frequency. The positive and negative frequency components actually just describe either the clockwise or the anti-clockwise rotation of the azimuthal mode $m=1 $.

From an engineering point of view, it is important to notice that all of the main structures detailed above have some pressure signature at the wall. Since we are only dealing with $m=1$ azimuthal Fourier mode, this means that they all contribute to the generation of side-forces. The features of lateral forces associated to the resonance mechanism are more precisely described in the following section. 

%%%%%%%%%%%%%%%%%%%%%%%%%%%%%%%%%%%%%%%%%%%%%%%%%%%%%%%%%%%%%%%
\section{Lateral pressure forces}
\label{lateral-pressure-forces}
\subsection{Description of wall pressure forces}
The aerodynamic forces resulting from the particular flow organization previously described are now examined based on the numerical data. 

\begin{figure}[htbp]
    \centering
    \includegraphics[width=\textwidth]{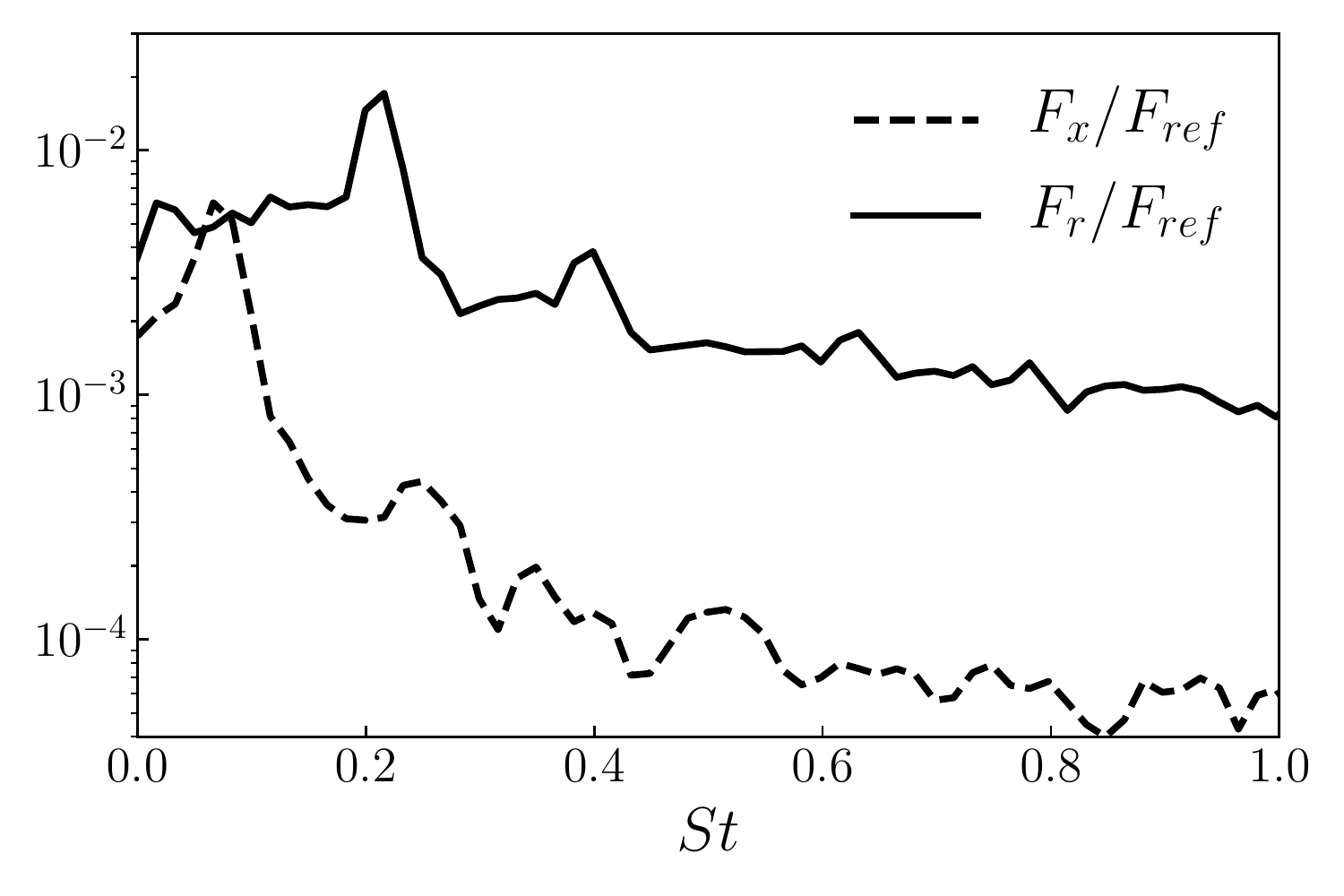}
    \caption{PSD of normalized aerodynamic forces acting on the nozzle in the axial direction $F_{x}$, and radial direction $F_{r}$.}
    \label{figure-side-load}
\end{figure}

The time evolution of the global force integrated all along the nozzle divergent is computed to examine the resulting spectral properties of each force component. 
The PSD of each force component is reported in Fig.~\ref{figure-side-load}.
The hump around $St=0.07$ in the PSD of $F_{x}$ can be related to the global axisymmetric motion of the upstream separation line. It should be reminded that rather low-frequency contributions to lateral forces are more often reported in other studies corresponding to separated jets without any clear tonal behaviour. Such low-frequency components are thus often associated only with dominant low-frequency shock motions for all modes. The most original observation in the present study is the emergence of the dominant peak also for lateral forces around $St \simeq 0.2$. This result thus contrasts with other results in the literature. Only moderate fluctuating energy levels of pressure mode $m=1$ are indeed observed locally within the flow. However, their coherence in the streamwise direction appears sufficiently important to lead to a dominant contribution to lateral forces. It should be noted that the amplitude of lateral forces at the tonal peak is even higher than the amplitude of thrust oscillations in the present case.

\subsection{Origins of lateral wall pressure forces}
Following \citet{Dumnov1996}, the side-loads Power Spectral Density $\mathbf{F_\omega}$ is the Fourier transform its time-domain autocorrelation function:
\begin{eqnarray}
\centering
\mathbf{F_\omega} &=& \int\!\!\!\!\int_{-\infty}^{\infty} 
    \mathbf{F}(t)\mathbf{F}^*(t+\tau) dt \, e^{-i\omega\tau} d\tau.
\end{eqnarray}
It is then straightforward to show that $\mathbf{F_\omega}$ is related to the first azimuthal mode of nozzle wall pressure:
\begin{eqnarray}
\mathbf{F_\omega} &=& 4\pi^2 \int_{0}^L \int_0^L  r(z) r(x) 
\mathbf{P}_{1,\omega}(x,z)dx \, dz,
\label{sl_psd}
\end{eqnarray}
where
\begin{eqnarray}
\mathbf{P_{1,\omega}}(x,z) & = & \int\!\!\!\!\int_{-\infty}^{\infty}
\mathtt{p}_1(z,t) \mathtt{p}^*_{1}(x,t+\
\tau) \, dt \, e^{-i\omega\tau} d\tau,
\end{eqnarray}
is the CSD of the antisymmetric azimuthal Fourier mode $m=1$ of the  wall pressure fluctuation $\mathtt{p}_1(x,t)$. As already mentioned by Dumnov \cite{Dumnov1996}, the side-loads PSD can only be obtained with a full knowledge of the wall pressure CSD. This two-point statistics can still only be obtained through numerical models.

\begin{figure}[htbp]
	\centering
	\includegraphics[width=0.8\textwidth]{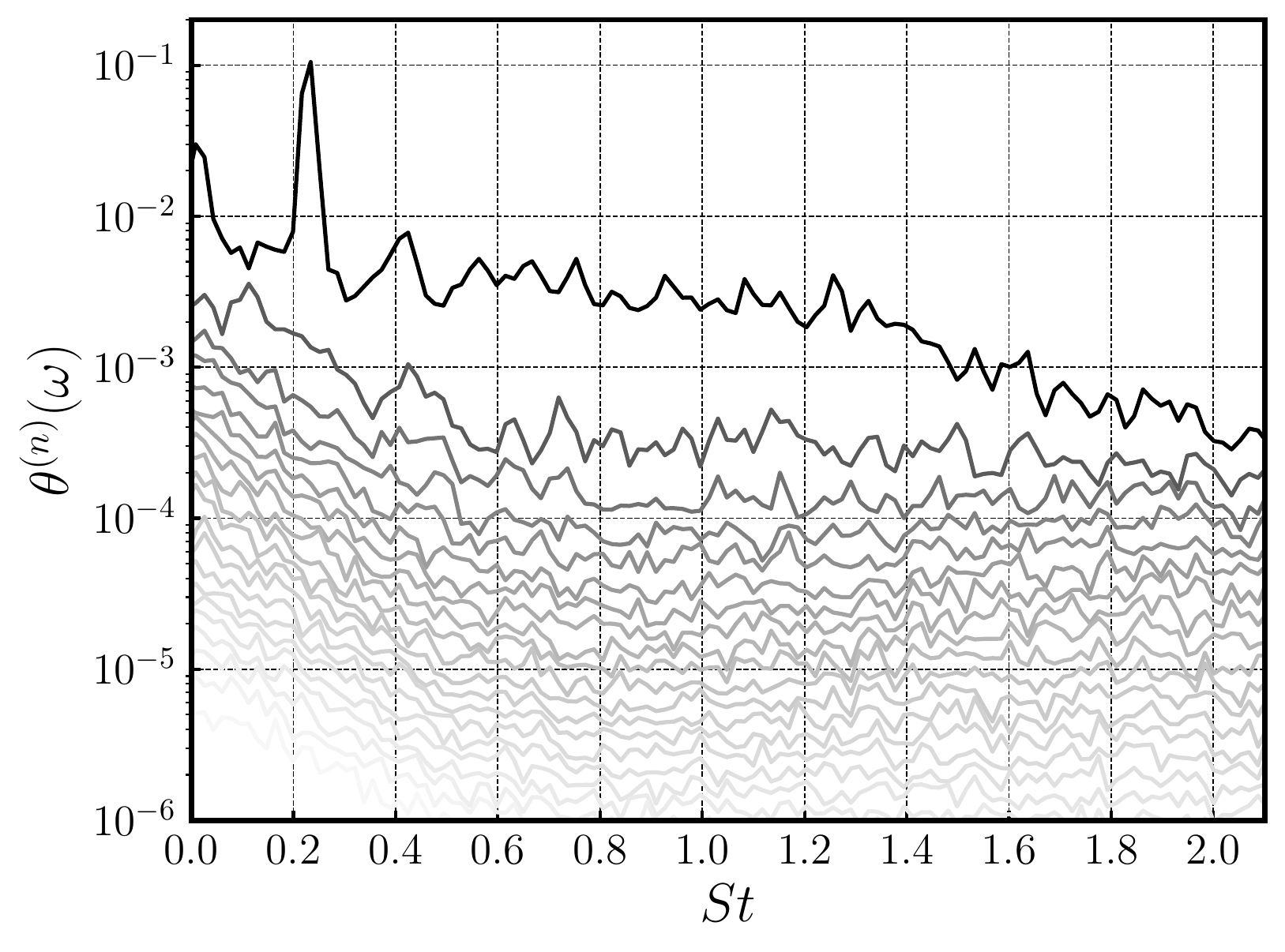}
	\caption{SPOD eigenvalues of the wall pressure signature as a function of Strouhal number.}
	\label{SPOD_wall_eigvalues}
\end{figure}

Taking advantage of the space-time discretisation of the domain, the side-loads formulae \ref{sl_psd} at the frequency $\omega$ can be written in matrix form as:
\begin{eqnarray}
  \mathbf{F_\omega} & = & \mathbf{w}^T \, \mathbf{P_{1,\omega}} \,\mathbf{w},
\end{eqnarray}
where $\mathbf{w}$ is a column vector of size corresponding to the number of discretization points at the wall, containing the integral weights: $\mathbf{w} = 2\pi L \cdot \mathbf{r}$. The CSD matrix of the antisymmetric wall pressure  is computed using Welch's periodogram technique:
\begin{eqnarray}
  \mathbf{P_{1,\omega}}&=& E\left\{ \hat{\mathrm{p}}_{1,\omega} \hat{\mathrm{p}}_{1,\omega}^H\right\},
\end{eqnarray}
where $E\left\{\cdot\right\}$ is the expectation operator. Similarly, to what has been done previously, the coherent structure acting on the wall can be obtained by decomposing the wall pressure CSD matrix into SPOD modes:
\begin{eqnarray}
  \mathbf{F_\omega} & = &  \mathbf{w}^T \; \mathbf{\Phi}_\omega \Theta_\omega \mathbf{\Phi}^H_\omega \; \mathbf{w},
\end{eqnarray}
where $\mathbf{\Phi}$ is the matrix formed whose columns are the SPOD modes vectors and $\Theta = diag(\theta)$ is the matrix of eigenvalues. This SPOD analysis only focuses on the wall pressure fluctuation, which is different from the previous analysis where we considered the entire fluid domain. There is no evidence, \textit{a priori}, that both analyses lead to identical features of coherent structures at the wall. 
This formulation is interesting because we can form a low rank side-loads amplitude by considering a truncation of the CSD using, for example, the N most energetic SPOD modes:
\begin{eqnarray}
  \mathbf{F_\omega} & \geq &  \mathbf{w}^T \left( 
  \sum_{i=1}^N \mathbf{\phi}^{(i)}_\omega 
  \theta^{(i)}_\omega
  \mathbf{\phi}^{(i)H}_{\omega} \right) 
  \; \mathbf{w}.
\end{eqnarray}
In case the wall pressure fluctuation possess a low rank dynamics, \textit{i.e} $\theta^{(1)} >> \theta^{(2)}$, the first SPOD mode of the wall pressure is responsible for the majority of the off-axis forces.\\
As for the analysis of the global field, the convergence of this SPOD analysis of wall fluctuating pressure field has been checked and similar observations are made. The eigenvalues are plotted in Fig.~\ref{SPOD_wall_eigvalues} as function of the Strouhal number. As can be seen, the first SPOD mode dominates the dynamics to the wall pressure fluctuations for a wide range of frequencies but even more at the resonance frequency $St\simeq 0.2$, where a difference of almost two orders of magnitudes can be found between the first and the second SPOD eigenvalues.

\begin{figure}[htbp]
	\centering
	    \subfloat[St=0.216]{
	    \includegraphics[width=0.8\textwidth]{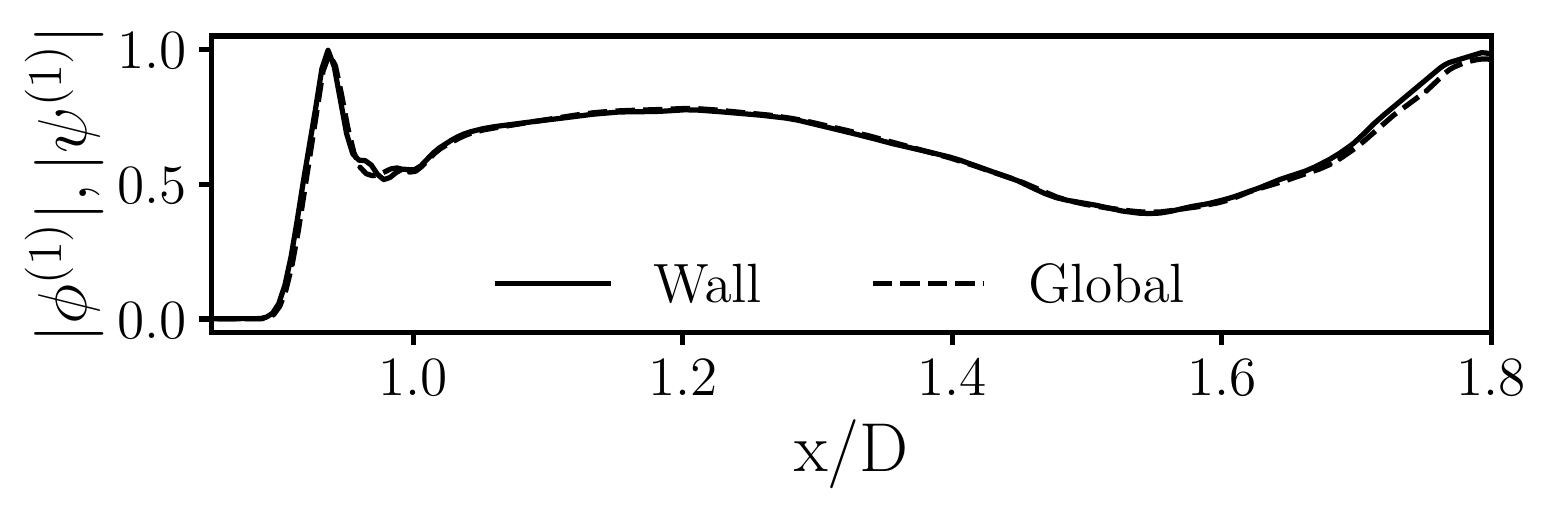}
	    }\\
	\caption{Comparison of wall pressure SPOD modes with the wall signature of global SPOD modes. The SPOD modes are normalized with respect to their maximum value in the plotted range.}
	\label{SPOD:spodmodes_global_wall_comp}
\end{figure}

The spatial distribution of the dominant SPOD mode of wall pressure fluctuations at the resonance frequency ($St=0.216$) is shown in Fig.~\ref{SPOD:spodmodes_global_wall_comp}. As expected, the structure of this SPOD mode is characterized by a peak at the separation shock foot ($x/D \simeq 0.9$). This is coherent with the early side-loads generation scenario, proposed by \citet{Schmucker1973-1}, involving asymmetric position of the separation point. From the results of Fig.~\ref{SPOD:spodmodes_global_wall_comp}, we can see however that in the separated region, downstream of the shock, the pressure fluctuations are also almost as high as the shock associated ones. This confirms, as proposed by \citet{Dumnov1996} and more recently by \citet{aghababaie2015modeling} in their dynamical model, the significant role of the separated region in the side-loads generation.\\
The wall pressure fluctuation associated with the global coherent structure found in the flow at the resonant frequency (see Fig.~\ref{SPOD:spodmodes_global}) is also reported in Fig.~\ref{SPOD:spodmodes_global_wall_comp}. It is striking that both the coherent wall pressure fluctuation and the wall pressure signature of the global coherent structure, identified earlier, collapse almost perfectly. This is an evidence that the global coherent structure, found in the jet flow, is responsible for most of the wall pressure fluctuations, and thus side-forces in the present case where a resonance emerges. From a general point of view, this shows that the shock motion and the separated region dynamics are not the only ingredients in the generation of side forces. It is clear from these results that the exiting jet flow dynamics also contributes and may not be neglected in side-loads modelling efforts.

\begin{figure}[htbp]
	\centering
	\includegraphics[width=0.8\textwidth]{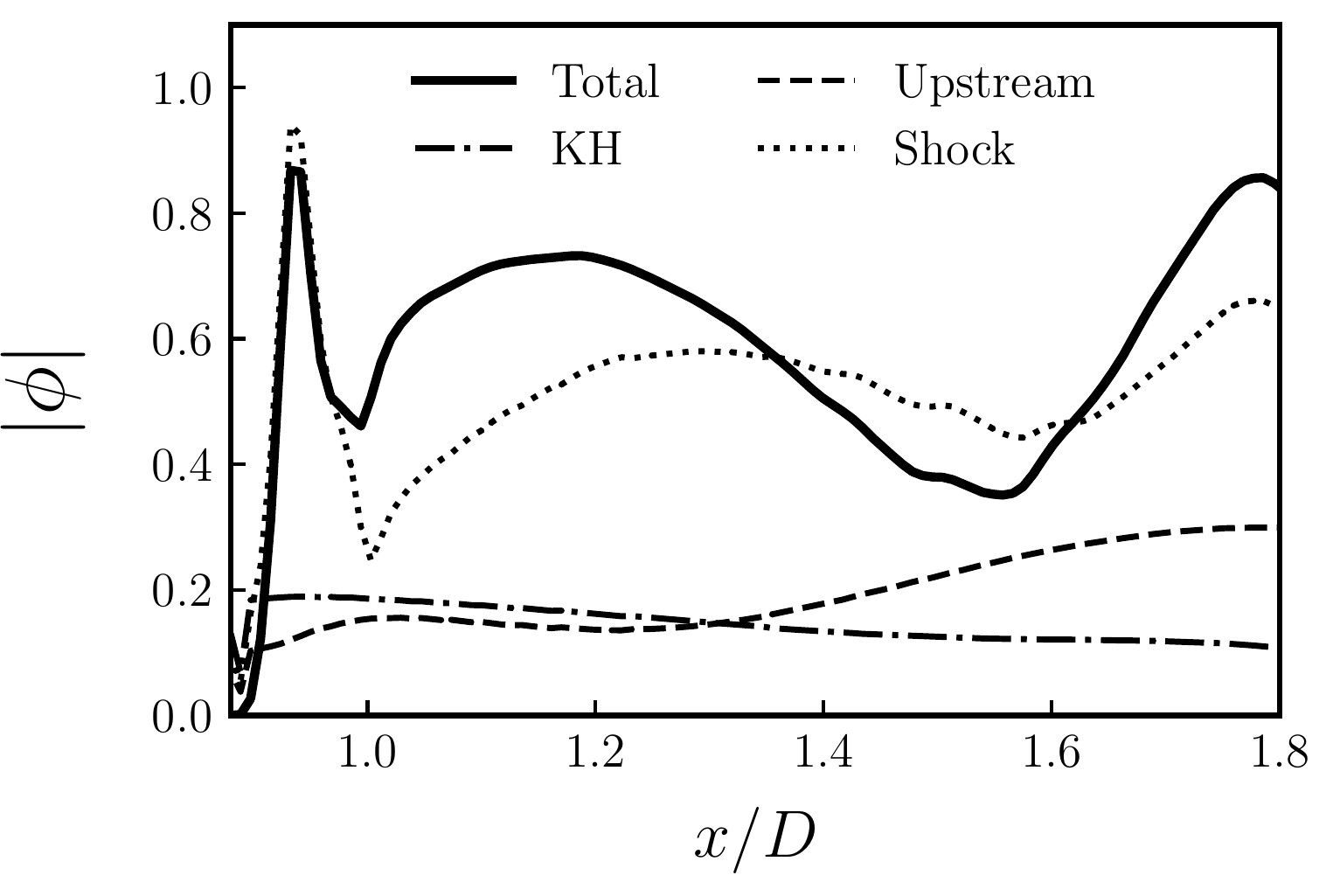}
	\caption{Comparison of wall pressure signature of the total SPOD mode with its individual components, filtered on spatial streamwise wavenumbers corresponding to downstream (KH) travelling waves, upstream travelling waves or remaining (Shock) contributions. The SPOD modes contributions are normalized with respect to the maximum value of the total SPOD mode in the plotted range.}
	\label{wall_pressure_contribution}
\end{figure}

We have evidenced that the jet flow coherent structure is responsible for the side-loads creation at the resonance frequency. Since we have previously shown that several distinct waves compose the structure of this coherent structure, \textit{i.e.} the KH wave, the external wave and the shock related structure, we would like to further investigate the role of each of these individual components onto the generation of side-loads. To this aim, the contribution  of each filtered dominant SPOD mode (see figures \ref{SPOD:spodmodes_filtered_recons_pos}, \ref{SPOD:spodmodes_filtered_recons_neg} and \ref{SPOD:spodmodes_filtered_recons_shock}) is computed:
\begin{eqnarray}
    \mathbf{\tilde{F}_\omega} & = &  \mathbf{w}^T \left( 
  \mathbf{\tilde{\psi}}^{(1)}_\omega 
  \theta^{(1)}_\omega
  {\mathbf{\tilde{\psi}}^{(1)}H}_{\omega} \right) 
  \; \mathbf{w}.
\end{eqnarray}
Indeed, the superposition of the individual contributions of the dominant SPOD mode filtered based on different spatial wavenumbers can be constructive or additive, \textit{i.e.} $\mathbf{F}_\omega \neq \sum \mathbf{\tilde{F}}_\omega$, nevertheless their relative importance can be of interest in aiming at understanding the side-loads generation mechanism. We present in Fig.~\ref{wall_pressure_contribution} the wall pressure signature of the downstream wave, the upstream one and the shock structure motion. It is clear that most of the wall pressure signature is related to the shock motion. This represents around 80\% of the sideloads amplitude at that frequency. The other two waves contributes less but still in a non-negligible manner: the downstream and the upstream waves represent around 7\%, and 13\% respectively, of the off-axis forces. 

Analysis of the shock related motion of the SPOD modes shows that the separation shock motion and the wall pressure in the whole separation region are indeed correlated. %An upstream motion of the shock with respect to its mean position is associated with an immediate increase of wall pressure, and a downstream motion of the shock is associated with a  decrease of wall pressure.
This behavior corroborates the propositions of \citet{Dumnov1996} regarding the main origins of side forces in separated nozzle flows. Our findings however suggest that 20\% of the side loads can be discarded when only the shock motion is accounted for.\\ 
The presented results are, to the authors opinion, strong indication that the resonance is formed via some sort of feedback loop involving an internal downstream propagating wave and an external upstream propagating one. The shock system reacts in order to adapt to these strong, resonating, pressure fluctuations, therefore creating most of the side forces. The exact nature of the waves at play in the resonance mechanisms and the identification of the conditions allowing their emergence deserve some attention in order to fully understand the origins of such a resonance. This issue should be addressed in future studies. %This is beyond the scope of this paper. 

%%%%%%%%%%%%%%%%%%%%%%%%%%%%%%%%%%%%%%%%%%%%%%%%%%%%%%%%%%%%%%%
\section{Conclusions and perspectives}
\label{conclusions}
The dynamical features of a separated supersonic jet in a TIC nozzle have been investigated for a nozzle pressure ratio corresponding to wall pressure oscillations of maximal amplitude. 
The wall pressure field shows signs of a particular tonal behaviour which has been previously reported. This study has focused on a new correlation analysis of synchronized time-resolved external velocity and internal wall pressure data to further understand the possible link existing between internal and external disturbances. In a narrow frequency band, a significant coherence level has been found between the antisymmetric internal wall pressure oscillations prevailing in the nozzle and antisymmetric external velocity disturbances spatially evolving in the jet mixing layer quite far downstream of the nozzle exit, underpinning the importance of the exiting jet plume dynamics in the nozzle wall pressure suggested in recent studies \cite{jaunet2017wall}. 
The flow behaviour has been numerically reproduced by DDES and validated with the available experimental results, allowing a detailed analysis of the resonant dynamics and an evaluation of the subsequent lateral forces applying to the nozzle. It has been shown, thanks to the use of a SPOD analysis, that the resonance is more likely due to a feedback loop involving a downstream-propagating and an upstream-propagating waves, similar to a screech resonance, but rather supported by the internal shear layer issued from the triple point. The present observations have not indicated any evidence of the need of instantaneous Mach disk curvature to sustain the feedback loop of the resonance, as recently proposed by \citet{martelli2019characterization}. The specific organization of the over-expanded mean flow, \textit{i.e.} an annular supersonic jet, makes the downstream waves mostly supported by the inner shear layer issued from the Mach disk triple point. This makes a major difference from the usual screech scenario, where both the upstream and downstream waves are supported by the same shear layer \cite{gojon2018oscillation,edgington2018upstream,mancinelli2019screech}. The present analysis could not indicate the reason for the occurrence of such resonance in over-expanded flows. Some research efforts need to be pursued to clarify the characterisitics of the waves involved and the end conditions that make them interact \cite{jordan2018jet,mancinelli2019screech}. This would allow to model and predict the occurrence of those undesirable tonal behavior.
Despite a relatively moderate local energy levels of the first azimuthal mode of wall pressure in this frequency range, it has been shown that these tonal jet oscillations fully dominate the whole dynamical behaviour of these lateral forces. The examination of the relative importance of the waves involved in the resonance in the generation of side-forces has shown that a non-negligible 20\% of the side forces can be attributed to the resonant waves. The remaining 80\% are shown to be attributed to the motion of the separation shock, but itself induced by the resonance.
These findings clearly suggest that the sole shock motion is insufficient to explain and model the side-forces amplitude in a general framework and that the jet flow dynamics should not be discarded in this respect. The results of the present study also indicate that any modification of the exiting jet environment is likely to affect not only the intrinsic external jet dynamics but also the side-loads features.

%%%%%%%%%%%%%%%%%%%%%%%%%%%%%%%%%%%%%%%%%%
%%%%%%%%%%%%%%%%%%%%%%%%%%%%%%%%%%%%%%%%%%
\section*{Acknowledgments}
Part of the present work has been carried out during the PhD Thesis of Florian Bakulu, supported by CNES and R\'egion Poitou-Charentes. This work was granted access to the HPC resources of TGCC under allocations A0012A10017 made by GENCI.

%%%%%%%%%%%%%%%%%%%%%%%%%%%%%%%%%%%%%%%%%%
\section*{References}
\bibliography{Jet-dance-TIC}

\end{document}